\documentclass[11pt]{article}

\usepackage{scalerel}
\usepackage{graphicx}
\usepackage{amsmath}
\usepackage{amsfonts}
\usepackage{amssymb}
\usepackage{dsfont}
\usepackage{physics}
\usepackage{microtype}
\usepackage{bbold}
\usepackage{bm}
\usepackage{mathtools}
\usepackage{cite}
\usepackage{cancel}
\usepackage{xcolor}
\usepackage{mdframed}
\usepackage{xpatch}
\usepackage{etoolbox}
\usepackage[normalem]{ulem}
\usepackage[colorlinks=true, linkcolor=blue, filecolor=blue, urlcolor=blue, citecolor=blue]{hyperref}
\usepackage[font=footnotesize,labelfont=bf,width=.9\textwidth]{caption}
\usepackage{mathtools}

\allowdisplaybreaks
\numberwithin{equation}{section}

\newcommand{\overbar}[1]{\mkern 1.5mu\overline{\mkern-1.5mu#1\mkern-1.5mu}\mkern 1.5mu}

\newcommand{\nn}{\nonumber} 
\renewcommand{\qq}{\qquad} 
\newcommand{\bn}{\bar{\nabla}}

\renewcommand{\title}[1]{\vbox{\center\LARGE{#1}}\vspace{5mm}}
\renewcommand{\author}[1]{\vbox{\center#1}\vspace{5mm}}
\newcommand{\address}[1]{\vbox{\center\footnotesize\em#1}}
\newcommand{\email}[1]{\vbox{\center\footnotesize\tt#1}\vspace{5mm}}

\newcommand{\bph}{\bm{\varphi}}
\newcommand{\bch}{\bm{\chi}}
\newcommand{\barc}{\bar{\chi}}
\newcommand{\barp}{\bar{\varphi}}
\newcommand{\Y}{\mathcal{Y}}

\newcommand{\vD}{\varphi}

\newcommand{\vonnt}{\tilde\varphi_{n}\tilde\varphi_{-n}}
\newcommand{\vot}{\tilde\varphi}
\newcommand{\vo}{\varphi}

\newcommand{\vdnnt}{\tilde\varphi_{n}\tilde\varphi_{-n}}
\newcommand{\vdt}{\tilde\varphi}
\newcommand{\vd}{\varphi}

\def\le{\left}
\def\ri{\right}

\begin{document}

\begin{titlepage}

\begin{center} 

\hfill \\
\hfill \\
\vskip 1cm

\title{The Effects of Near-AdS$_2$ Backreaction on Matter Fields}

\vskip10mm

\author{Alejandra Castro$^{a}$, Jildou Hollander$^{b}$, Pedro J. Martinez$^{c}$ and Evita Verheijden$^{d,e,f,g}$
}

\vskip1em

\address{  $^{a}$Department of Applied Mathematics and Theoretical Physics,  University of Cambridge, \\
Cambridge CB3 0WA, UK\\ 
\vspace{0.7em}

$^{b}$Institute for Theoretical Physics, University of Amsterdam,  1090 GL Amsterdam, The Netherlands\\

\vspace{0.7em}

\it $^{c}$ 
Instituto de Física La Plata - CONICET, Universidad Nacional de La Plata, \\
La Plata, C.C. 67, 1900, Argentina\\

\vspace{0.7em}

$^{d}$Black Hole Initiative at Harvard University,
20 Garden Street, Cambridge, MA 02138, USA\\

\vspace{0.7em}

$^{e}$Center for Theoretical Physics -- a Leinweber Institute, MIT, Cambridge, MA 02139, USA\\

\vspace{0.7em}

$^{f}$Nordita, KTH Royal Institute of Technology and Stockholm University, 106 91 Stockholm, Sweden \\

\vspace{0.7em}

$^{g}$The Oscar Klein Centre and Department of Physics,
Stockholm University, 106 91 Stockholm, Sweden
}
\vskip10mm

\email{ac2553@cam.ac.uk, j.s.hollander@uva.nl, \\ martinezp@fisica.unlp.edu.ar, evita.verheijden@fysik.su.se}
\end{center}

\begin{abstract}
We quantify how the two-point function of a real scalar field is affected by the distortion caused by deforming AdS$_2$ to a near-AdS$_2$ background. At tree-level, the backreaction of the geometry induces a finite-temperature correction to the correlator that arises from interactions that the background generates. For a massive field, we show that this correction is not captured by JT gravity coupled to matter: it requires a backreaction of the metric field. For a massless field, the correction is controlled solely by the dilaton and hence is model-independent. 
We compare our findings with correlation functions on BTZ and find perfect agreement. We use our results to quantify the corrections for a class of correlators relevant to five- and four-dimensional black holes. We discuss how these corrections would enter in a holographic description of near-AdS$_2$; we also comment on how these corrections provide a universal prediction for quasinormal modes in higher dimensions. 
\end{abstract}

\end{titlepage}
 
\tableofcontents
\newpage

\section{Introduction}

Two-dimensional theories of dilaton gravity have provided valuable insights about both classical and quantum gravity. A prominent example is that of Jackiw--Teitelboim (JT) gravity, which is a simple model of a real scalar field, a.k.a. the dilaton, coupled to AdS$_2$ gravity \cite{Jackiw:1984je,Teitelboim:1983ux}. The reason for its prominence is two-fold: this theory universally encodes the leading gravitational backreaction away from an idealised AdS$_2$ geometry \cite{AlmPol14,MalSta16}, which is essential to describe properties of near-extremal black holes, and it is a solvable model of quantum gravity \cite{SaaShe19}.  

In recent years, there has been a renewed interest in making the connection between JT gravity and higher-dimensional near-extremal black holes more precise. One approach to capture the influence of JT gravity on black holes is to inspect the dynamics of near-extremal black holes: either from the equations of motion \cite{Porfyriadis:2018jlw,HadPod20,CasGod21,MarTol25,CasMan25}, or from the gravitational path integral \cite{IliTur20,IliMur22,KapShe23,RakRan23,BanSah23,KapLaw24,KolMar24,ArnBon24}. Another approach is to consistently account for the degrees of freedom that arise from the higher-dimensional theory in the two-dimensional effective description. Examples along these lines include the effective theories constructed in \cite{CvePap16,Lar18,CasLar18,CasVer21}. Here, we will take the latter approach and systematically investigate these two-dimensional effective theories. In particular, we will quantify how the backreaction of AdS$_2$ affects the dynamics of the matter fields appearing in such models. 
 
The two-dimensional models that arise from a consistent dimensional reduction incorporate features that cannot be described by simply adding matter fields to JT gravity. For this reason, we will start with a more general class of dilaton gravity models: two-dimensional theories with a general potential for the dilaton and including an external scalar field coupled to both the metric and dilaton. The models are designed to contain AdS$_2$ as a background solution, and will have the distinctive backreaction of \cite{AlmPol14,MalSta16}.
Unlike in JT gravity, in the models we consider, the metric will also backreact due to the presence of the dilaton. This is essential to describe the effects of the backreaction on the scalar field, and hence make the appropriate predictions.  

For these general dilaton gravity models, we will describe the resulting effective action for matter fields, focusing on the interactions between the scalar field and the backreacted metric and dilaton. At tree-level, these interactions are finite-temperature corrections to correlation functions, as we will show in detail for two-point functions. Na\"ively, one might suspect that a minimal coupling between the dilaton and matter fields suffices as a toy model to describe this temperature correction. By exploiting the gauge freedom in two dimensions, we will show that this is incorrect, and interestingly, the result depends on the mass of the scalar field. For a massive field, a cubic interaction between the dilaton and the field is pure gauge; the interactions with the metric contain all the information related to the backreaction (hence it is key that the metric field backreacts). For a massless field, the metric has no minimal coupling with the field, and the backreaction of these fields is therefore universal in that it only depends on the coupling to the dilaton field. With the resulting effective actions at hand, we will evaluate the leading correction to the two-point function of the scalar fields due to the backreaction of AdS$_2$ for massless fields and massive fields with integer conformal dimension. This analysis rectifies the correction reported in \cite{MalSta16}, which we explain in the main sections. 

To test our findings, we turn to a simple black hole setting where we have control: the three-dimensional BTZ black hole. This is a situation where the two-point function of a scalar operator on a fixed black hole background is known at finite temperature. Here we will see that the correction due to the backreaction of AdS$_2$ precisely matches the low-temperature corrections of the correlation function on BTZ. This confirms that our treatment of the interactions in two dimensions is indeed sound. In addition, since the analysis of two-point functions on BTZ is also valid for non-integer conformal dimensions, this match emboldens us to predict a general form for the correction for generic values of the conformal dimension. 

We will apply our findings on the corrected two-point function to two different dilaton gravity models that arise from a dimensional reduction: four-dimensional ${\cal N} = 2$ $U(1)^4$ supergravity reduced on a two-sphere \cite{CasVer21}, and the five-dimensional Einstein--Hilbert action with a negative cosmological constant reduced on a squashed three-sphere \cite{CasLar18}. In each of these cases, there is a distinct potential for the dilaton, which will have an imprint on the backreaction of the metric and, consequently, influence the magnitude of the correction. Another appeal of these models is that the matter content arises naturally from the properties of the black hole solutions underlying the theory.  There is also a wide spectrum of masses for the matter fields, which is a feature controlled by the details of the surroundings and background in higher dimensions: (1) whether the theory is asymptotically flat or AdS, and (2) whether the AdS$_2$ background preserves supersymmetry or not. 

These two models are also of interest since the corrections to the two-point function were reported in \cite{CasPed21,CasVer21}, following the prescription in \cite{MalSta16}, and they led to some puzzling aspects. For the 5D example, \cite{CasPed21} found that the cubic couplings among the squashing mode and the dilaton field did not have a definite sign; this is puzzling from the point of view of the putative dual theory, in which such interactions are related to a correction to the density of states, and should therefore be positive-definite. For the 4D black holes studied in \cite{CasVer21}, fluctuations around non-BPS backgrounds lead to extremal correlators between the dilaton field and a massless scalar field present in the theory, which are pathological. 

Here we find the root cause of both of these puzzles: in computing the imprint on the two-point function of the scalar fields due to the cubic interactions, we treated the background value of the dilaton as an operator with $\Delta = -1$ and static source, as advocated first by \cite{MalSta16}. However, in doing so, one makes two mistakes: first, for massive scalar fields, which are relevant for the 5D rotating black hole in \cite{CasPed21}, we did not identify correctly how the backreaction affects the scalar field; the vertex considered in \cite{MalSta16} and used also in \cite{CasPed21} is actually pure gauge and can be removed, as we show here. Second, for massless fields, relevant to the theories from 4D SUGRA, treating the dilaton as a fluctuating field with $\Delta = -1$ yields the wrong coefficient for the correction to the two-point function. With the correct coefficient, the divergence of the extremal correlator actually is cancelled. By carefully evaluating the relevant vertex, which differs significantly between massive and massless fields, we can give an explicit expression for the full leading order tree-level correction to the two-point functions of matter fields due to their interactions with the dilaton field and the graviton (which at tree-level is sourced by the dilaton field), for integer conformal dimensions. 

Finally, it is important to emphasise that our computations are classical, as we only treat interactions at tree-level. Much attention has been given in recent years to the quantum backreaction of AdS$_2$, which is described by the Schwarzian action \cite{ChaLar19,IliTur20}. Although there are certainly works that discuss the imprint of this quantum backreaction on matter fields present in the theory---notably, to compute the quantum cross section of near-extremal black holes \cite{Emp25,BetPap25} and the evaporation of black holes in the quantum regime \cite{BroIli24, Big25}---they do not account for the classical backreaction of the metric, thereby evading any of the effects that we describe here. A full account of the imprint of near-AdS$_2$ on matter fields requires both the classical and quantum backreaction. Our goal here is to explore the classical backreaction and describe the tree-level effects.

This paper is organised as follows. In Sec.\,\ref{sec:general-2d-EA}, we describe the general class of dilaton gravity theories coupled to matter that we will consider. We describe the AdS$_2$ background, the linearised fluctuations around it, and the effective theory that describes the backreaction. We then discuss the effect of the near-extremal backreaction on the matter fields, distinguishing between massive and massless fields. In Sec.\,\ref{sec:corr-2pt-fn} we quantify precisely how the backreaction of AdS$_2$ affects the correlation functions of the operators ${\cal O}_\Delta$ dual to the scalar field of mass $\mathfrak{m}^2 \ell_2^2 = \Delta(\Delta-1)$. We explicitly evaluate the interaction vertex for $\Delta = 1$ (massless) and $\Delta = 2$, and then give the general form for the corrected two-point function of ${\cal O}_\Delta$ for any positive integer $\Delta$. In Sec.\,\ref{sec:BTZ-BHs}, we compare these results with the near-extremal limit of a correlation function of scalar operators on the BTZ black hole geometry, achieving a perfect match for integer $\Delta$. This analysis will also provide a prediction for the near-AdS$_2$ backreacted correlators for generic values of the conformal dimensions. In Sec.\,\ref{sec:higher-d-exs}, we discuss our results in the context of two higher-dimensional examples: five-dimensional rotating black holes in AdS or flat space, and non-BPS dyonic black holes in ${\cal N}=2$ 4D $U(1)^4$ ungauged supergravity. We solve some puzzling features that arose in similar studies of these theories in earlier work. We end in Sec.\,\ref{sec:discussion} by a careful discussion of our results, and comment on potential future directions. We also include two appendices. In App.~\ref{App:Vertex}, we contrast our methods and results with those in \cite{MalSta16} regarding the coupling between matter and the dilaton; and in App.~\ref{app:rec-rel}, we provide details of the derivations in Sec.\,\ref{sec:massive-explicit-vertex}.

\section{Coupling matter to dilaton gravity} \label{sec:general-2d-EA}
 
In this section, we will consider a class of two-dimensional dilaton gravity models, equipped with a metric $g_{ab}$ and a dilaton $\Phi$, coupled to a massive (or massless) scalar field $\varphi$. The case of interest is when this theory contains AdS$_2$ as a background solution; from this vacuum, we will perturb the system and analyse its backreaction. The dilaton drives the response to leading order, and here we will carefully construct the appropriate effective theory to see the effects of the backreaction on the dynamics of $\varphi$.     

\subsection{Effective action near-AdS\texorpdfstring{$_2$}{2}}\label{sec:gen-bg-eqs}

Inspired by the effective theories that we will study in the following sections, a simple model that will capture the appropriate complexities is
    \begin{equation} \label{eq:gen-dil-model}
        S_{\rm EFT} = \frac{1}{2\kappa_2^2} \int \dd^2x \sqrt{-g}\,  \left( \Phi \, R+ V(\Phi) - \frac{1}{2}\Phi\, \partial_a\varphi\partial^a\varphi - \frac{\mathfrak{m}^2}{2} \Phi\, \varphi^2 \right).
    \end{equation}
Here $R$ is the two-dimensional Ricci scalar, $\kappa_2^2$ is the dimensionless Newton's constant in two dimensions, and $\mathfrak{m}^2$ is the mass parameter for $\varphi$. For the moment, we allow for a general potential $V(\Phi)$. We have made two important simplifications in writing \eqref{eq:gen-dil-model}. First, the scalar $\varphi$ only enters quadratically in the action. Our aim is to quantify how the backreaction of $\Phi$ affects the free propagation of $\varphi$, and self-interactions are a subleading effect. 
Second, we could have added different powers of $\Phi$ for how the dilaton couples to the  Ricci scalar relative to kinetic and to mass terms. Since we will only look at the leading effect of the backreaction, it suffices to keep all the powers linear.   

We begin by discussing the relevant AdS$_2$ background, and then consider linearised perturbations around this solution; here we identify the role of the dilaton on the backreaction and its imprints on the metric field. Finally, we construct the effective action that captures the effect of this backreaction on the scalar field, which results in a modified cubic interaction relative to the na\"ive couplings in the full action \eqref{eq:gen-dil-model}.   
    
\paragraph{Background AdS$_2$ solution.} In this portion, we will construct the vacuum solution $(\varphi_0, \Phi_0, \bar g_{ab})$ around which we will expand the action. The equations of motion for the action \eqref{eq:gen-dil-model} read
    \begin{subequations}\label{eq:all-eom}
    \begin{align}
        \nabla_a \left(\Phi\, \nabla^a\varphi\right) - \mathfrak{m}^2\,\Phi\, \varphi = 0~,\label{eq:eom-varphi}\\
         R + V'(\Phi) - \frac{1}{2}\big( (\partial \varphi)^2 + \mathfrak{m}^2\varphi^2\big) =0~,\label{eq:eom-phi}\\
        (\nabla_a \nabla_b - g_{ab} \square)\Phi + \frac{1}{2}g_{ab} V(\Phi) + \frac{1}{2}\Phi  \nabla_a\varphi\nabla_b \varphi- \frac{1}{4} g_{ab}\Phi\big((\partial\varphi)^2 + \mathfrak{m}^2\varphi^2\big)  = 0~.\label{eq:eom-metric}
        \end{align}
    \end{subequations}
We are seeking a solution for which the scalars $(\varphi_0,\Phi_0)$ are constants. From \eqref{eq:eom-varphi}, we immediately get\footnote{If the field is massless, we just have $\varphi_0={\rm constant}$. The rest of the discussion is unchanged.}
\begin{equation}\label{eq:vacua-varphi}
    \varphi_0=0~.
\end{equation}
The remaining two equations of motion are then simplified to 
 \begin{equation}\label{eq:eom-phi-vac}
         \bar R + V'(\Phi_0) =0~,\qquad V(\Phi_0)= 0~.
    \end{equation}
This tells us that $\Phi_0$ is a zero of the potential, and since we are interested in having AdS$_2$ vacua, we will require that $V'(\Phi_0)>0$ and therefore define the AdS$_2$ radius $\ell_2$ as
\begin{equation}\label{eq:eom-ads2}
    \bar R = -  V'(\Phi_0) =: - \frac{2}{\ell_2^2}~.
\end{equation}

\paragraph{Linear analysis.}\label{sec:gen-lin-analysis}
We now discuss the linear fluctuations around the AdS$_2$ background. To this end, we define
    \begin{equation}
    \begin{aligned}
 \varphi&= \varphi_0  +\epsilon \,\hat \varphi\,\\   
        \Phi &= \Phi_0 + \epsilon \,\Y\,, \\
        g_{ab} &= \bar{g}_{ab} + \epsilon\, h_{ab}\,,
    \end{aligned}
    \label{eq:lin-fluc}
    \end{equation}
where $(\varphi_0,\Phi_0, \bar{g}_{ab})$ are the background values satisfying \eqref{eq:vacua-varphi}-\eqref{eq:eom-phi-vac}. We introduced a dimensionless parameter $\epsilon$ to control how much we deviate away from the AdS$_2$ vacua, i.e., the strength of the backreaction. 
To linear order in $\epsilon$, the equation of motion for the scalar field is
    \begin{equation} \label{eq:gen-lin-scalar-eq}
        \bar\square \hat \varphi =  \mathfrak{m}^2 \hat \varphi~,
    \end{equation}
which simply tells us that to leading order $\hat \varphi$ is a field of mass $\mathfrak{m}^2$ propagating on AdS$_2$. Its dual conformal dimension is $\Delta(\Delta-1)=\ell_2^2\mathfrak{m}^2$.

The Einstein equation \eqref{eq:eom-metric}, again to linear order in $\epsilon$, is 
    \begin{equation}\label{eq:Y-eom}
         (\bn_a \bn_b - \bar{g}_{ab}\bar\square)\Y - \frac{\bar{R}}{2} \bar{g}_{ab}\Y = 0~,
    \end{equation}
which shows that $\Y$ satisfies the equation of motion of the dilaton in JT gravity. Upon using $\bar{R} = - \frac{2}{\ell_2^2}$, the trace of this equation is
    \begin{equation}\label{eq:eom-boxY}
        \bar\square\Y = \frac{2}{\ell_2^2} \Y \,,
    \end{equation}
which mimics a field of mass $\ell_2^2 \mathfrak{m}^2 = 2$. However, the dynamic of $\Y$ is rather special. It is not on the same footing as $\hat{\varphi}$, for two reasons. First, it is heavily constrained by \eqref{eq:Y-eom}, where the solutions are simple functions of the coordinates of AdS$_2$. Hence $\Y$ does not propagate freely on AdS$_2$ in contrast to $\hat{\varphi}$. Second, any addition of energy to the AdS$_2$ background requires $\Y$ to be non-trivial with a source turned on \cite{AlmPol14}.\footnote{See \cite{MerTur22} for a detailed derivation of this argument.} For this reason, the backreaction of AdS$_2$ is driven by $\Y$. 

Finally, the dilaton equation \eqref{eq:eom-phi} gives at linear order 
    \begin{equation} \label{eq:lin-dil}
        \bar{R}_{ab} h^{ab} - \bn^a\bn^b h_{ab} + \bar\square h^a_{~a} + \frac{\nu}{\ell_2^2} \Y = 0~,
    \end{equation}
where we defined 
    \begin{equation} \label{eq:ch-eq}
        \frac{\nu}{\ell_2^2} :=  - V''(\Phi_0)~.
    \end{equation}
This last equation for $h_{ab}$ is rather important: it tells us that there is a backreaction of the metric due to $\cal Y$. This is not a feature of JT gravity, where the metric remains unchanged as $\cal Y$ is turned on. As we will see, the inhomogeneous solution to \eqref{eq:lin-dil} will be important in extracting the correct cubic vertex that couples $\hat{\varphi}$ to $\cal Y$.  With this, we define the \textbf{near-AdS$_2$} background as the solution with matter fields turned off and
  \begin{equation}
    \begin{aligned}
        \Phi &= \Phi_0 + \epsilon \,\Y\, + \cdots, \\
        g_{ab} &= \bar{g}_{ab} + \epsilon\, h_{ab}(\Y)+ \cdots\,,
    \end{aligned}
    \label{eq:near-AdS2}
    \end{equation}
where $h_{ab}(\Y)$ is a solution to \eqref{eq:lin-dil}. The dots in this equation represent higher powers of $\Y$, which are subleading corrections for the purpose of our analysis. 

\paragraph{Interactions near AdS$_2$.} \label{sec:gen-cubic-action}
Next, we investigate the response to the system by adding subleading terms in $\epsilon$, where $\epsilon$ is the parameter introduced in \eqref{eq:lin-fluc}. Expanding the action \eqref{eq:gen-dil-model} up to order $\epsilon^3$ around the AdS$_2$ background, we will have three contributions 
\begin{equation}
    S_{\rm eff} = \int d^2 x \sqrt{-\bar{g}} \left( {\cal L}_{\rm kin} + {\cal L}_{\rm int} + {\cal L}_{\rm grav}\right) + O(\epsilon^4)~. 
\end{equation}
Here ${\cal L}_{\rm kin}$ are the kinetic terms for $\hat\varphi$ appearing at order $\epsilon^2$. Explicitly we have
\begin{equation}
    {\cal L}_{\rm kin} = \frac{\epsilon^2}{2\kappa_2^2} \Phi_0\left(-\frac{1}{2}(\partial \hat \varphi)^2 - \frac{1}{2} \mathfrak{m}^2\hat \varphi^2 \right)~.
\end{equation}
We can normalise $\hat\varphi$ to absorb the prefactor of $\Phi_0\epsilon^2/(2\kappa_2^2)$, and with a slight abuse of notation, rename the field $\varphi$. This gives
\begin{equation}\label{eq:free-scalar}
    {\cal L}_{\rm kin} = -\frac{1}{2}(\partial \varphi)^2 - \frac{1}{2} \mathfrak{m}^2 \varphi^2 ~,
\end{equation}
where $\varphi = \sqrt{\Phi_0\epsilon^2/(2\kappa_2^2)}\hat\varphi$.
The term ${\cal L}_{\rm int} $ is of order $\epsilon^3$, containing terms of the form $\Y \hat \varphi^2$ and $h \hat \varphi^2$. In terms of the rescaled field, $\hat \varphi\to \varphi$, we have
\begin{equation}\label{eq:int-scalar-Y}
    {\cal L}_{\rm int} =  - \frac{\epsilon}{2\Phi_0} \left(\Y(\partial \varphi)^2 +\mathfrak{m}^2 \Y \varphi^2  +  \frac{\Phi_0}{2}h^a_{~a} \left(  (\partial\varphi)^2 + \mathfrak{m}^2\varphi^2\right) - \Phi_0 h^{ab} \partial_a \varphi \partial_b \varphi \right).
\end{equation}
It is worth noting that the interactions controlled by $\Y$ are suppressed by $\Phi_0$.\footnote{Thus, to leading order in the large $\Phi_0$ limit, the matter fields do not couple directly to the dilaton field $\Y$. This is important when studying quantum effects.} From a higher-dimensional perspective, $\Phi_0$ is the area of the extremal horizon in units of the volume of the compactification. Note that \textit{prima facie} the interactions controlled by $h_{ab}$ are not suppressed by $\Phi_0$. However, recall that we will work at tree-level, where $h_{ab} = h_{ab}(\Y)$; we will see that solving $h_{ab}$ in terms of $\Y$ introduces a factor $\nu$, which will generically scale as $\Phi_0^{-1}$. 

Finally, we have the terms intrinsic to the gravitational sector, which couple $\Y$ with the metric perturbation $h$:
   \begin{equation} \label{eq:L-int-hab-Y}
    \begin{aligned}
        \mathcal{L}_{\rm grav} =& \frac{\epsilon^3}{8\kappa_2^2} \Big( 2\Y h^{ab} \bn_a \bn_b h^c_{~c} + 2\Y h^{ab} \bar{\square}h_{ab} - 4 \Y h^{ab} \bn_a \bn_c h_b^{~c} \\
         &- 2\Y (\bn^a h_{ab})(\bn^c h^b_{~c}) + \Y (\bn^a h^b_{~b})(\bn_a h^c_{~c}) + \Y (\bn^a h_{bc})(\bn_a h^{bc}) \Big).
    \end{aligned}
    \end{equation}
These terms control the graviton propagator \cite{AlmKan16}, which is important when evaluating quantum corrections that involve the exchange of gravitons (see also \cite{CasPed21} for further technical details). To incorporate these effects one would then consider, schematically,
    \begin{equation}
        h_{ab} = h_{ab} (\Y) +h_{ab}^{\scaleto{\rm quantum}{4pt}}~,
    \end{equation}
separating the near-AdS$_2$ backreaction from the quantum corrections. 
Our forthcoming analysis will only include the former (tree-level) effects; the quantum corrections, and therefore \eqref{eq:L-int-hab-Y}, will not play a role. 
 
\subsection{Backreaction of \texorpdfstring{$\Y$}{Y} on matter fields} \label{sec:gen-corr-2pt}

The dilaton field $\Y$ plays a central role in deviating the system away from the pristine AdS$_2$ background. In that sense, as we study how matter fields are affected by the backreaction of the AdS$_2$, we should be evaluating the effective action of $\varphi$, given by \eqref{eq:free-scalar} plus \eqref{eq:int-scalar-Y}, on the near-AdS$_2$ background \eqref{eq:near-AdS2} to capture the leading order effect. The purpose of this subsection is to capture explicitly how $h_{ab}$ depends on $\Y$, and write a more practical version of the Lagrangian \eqref{eq:int-scalar-Y}.

We start by discussing the solutions to \eqref{eq:lin-dil}, which reads
    \begin{equation} \label{eq:lin-dil-2}
         \bar\square h^a_{~a}- \frac{1}{\ell_2^2}h^a_{~a} - \bn^a\bn^b h_{ab} = -\frac{\nu}{\ell_2^2} \Y ~.
    \end{equation}
Here we used that $\bar R_{ab}=-\bar g_{ab}/\ell_2^2 $. It is useful to split the solutions into homogeneous and inhomogeneous terms: $h_{ab}=h^{\scaleto{\rm hom}{4pt}}_{ab}+h^{\scaleto{\rm inh}{4pt}}_{ab}$. The homogeneous solution to this equation is a diffeomorphism acting on the background metric, i.e.,
    \begin{equation}\label{eq:h-diffeo}
        h^{\scaleto{\rm hom}{4pt}}_{ab}= {\cal L}_\zeta \bar g_{ab}= \bn_a \zeta_b + \bn_b \zeta_a~.
    \end{equation}
One particularly useful diffeomorphism is to choose $\zeta_a= \frac{1}{2}\alpha \partial_a \Y $, for which we would have 
    \begin{equation}\label{eq:h-hom}
         h^{\scaleto{\rm hom}{4pt}}_{ab}= \frac{\alpha}{\ell_2^2} \bar g_{ab} \Y~,
    \end{equation}
where $\alpha$ is an arbitrary constant. This homogeneous solution is quite important in simplifying the terms that appear in \eqref{eq:int-scalar-Y}. To see this, we first replace \eqref{eq:h-hom} in ${\cal L}_{\rm int}$; when $\mathfrak{m}^2\neq 0$, this gives\footnote{It is worth mentioning that the interaction Lagrangian \eqref{eq:int-scalar-Y} is gauge invariant. This implies that if $h_{ab}$ is of the form \eqref{eq:h-diffeo}, then the corresponding terms in \eqref{eq:int-scalar-gauge-2} lead to a total derivative to leading order in $\epsilon$. This might seem in tension with \eqref{eq:int-scalar-Y-1}, but there is no contradiction. If one uses that $\bn_a \bn_b \Y = \frac{1}{\ell_2^2} g_{ab}\Y$, then \eqref{eq:int-scalar-Y-1} is a total derivative in disguise.}
\begin{equation}\label{eq:int-scalar-Y-1} 
\begin{aligned}
      \frac{1}{2}(h^{\scaleto{\rm hom}{4pt}})^a_{~a} \left(  (\partial\varphi)^2 + \mathfrak{m}^2\varphi^2\right) - (h^{\scaleto{\rm hom}{4pt}})^{ab} \partial_a \varphi \partial_b \varphi 
      = \frac{\alpha}{\ell_2^2} \,\mathfrak{m}^2\, \Y\, \varphi^2 ~.
\end{aligned}
\end{equation}
This shows that for massive fields any term proportional to $\Y\varphi^2$ is pure gauge. Next, we can use
 \begin{equation}
 \begin{aligned}\label{eq:id-int}
        \sqrt{-\bar g}\,\Y (\partial\varphi)^2 &= {\rm tot.~der.~} + \frac{1}{2} \sqrt{-\bar g}\,\varphi^2 \bar \Box \Y  -  \sqrt{-\bar g}\,\Y \varphi \bar \Box \varphi \\ 
        &= {\rm tot.~der.~} + \frac{1}{\ell_2^2}  \sqrt{-\bar g}\, \Y \varphi^2 -\mathfrak{m}^2  \sqrt{-\bar g}\, \Y \varphi^2 + O(\epsilon)~,
 \end{aligned}
   \end{equation}
where in the second line we used the linearised equations \eqref{eq:gen-lin-scalar-eq} and \eqref{eq:eom-boxY}. Combining \eqref{eq:int-scalar-Y-1} and \eqref{eq:id-int} in \eqref{eq:int-scalar-Y}, we find
\begin{equation}\label{eq:int-scalar-gauge-2}
\begin{aligned}
    {\cal L}_{\rm int} =&  - \frac{\epsilon}{2\Phi_0\ell_2^2} \left( \Y \varphi^2  +  \alpha  \Phi_0 \,\mathfrak{m}^2\, \Y\, \varphi^2  \right)\\ &~~- \frac{\epsilon}{4}(h^{\scaleto{\rm inh}{4pt}})^a_{~a} \left(  (\partial\varphi)^2 + \mathfrak{m}^2\varphi^2\right) + \frac{\epsilon}{2} (h^{\scaleto{\rm inh}{4pt}})^{ab} \partial_a \varphi \partial_b \varphi + O(\epsilon^2)~. 
\end{aligned}
\end{equation}
Thus, for massive fields, the interaction Lagrangian reduces to the second line of \eqref{eq:int-scalar-gauge-2}, since in this case the terms proportional to $\Y\varphi^2$ are pure gauge. In general, any term of the form $\Y \varphi^2$ and $\Y (\partial \varphi)^2$ can be removed from the action if $\mathfrak{m}^2\neq0$, to leading order in $\epsilon$. In the following, we will discard any such terms in the interactions. This also brings a dichotomy between massive and massless fields, which we now discuss separately.  

\paragraph{Massless fields ($\mathfrak{m}^2 = 0$).} In this case, introducing the homogeneous solution \eqref{eq:h-hom} has no effect. Still, there are some important simplifications in this case due to the field being massless. From \eqref{eq:int-scalar-gauge-2} we have
\begin{equation}\label{eq:int-mass}
\begin{aligned}
    ({\cal L}_{\rm int})_{\mathfrak{m}^2=0} =&  - \frac{\epsilon}{2\Phi_0\ell_2^2}  \Y \varphi^2  - \frac{\epsilon}{4}(h^{\scaleto{\rm inh}{4pt}})^a_{~a} (\partial\varphi)^2  + \frac{\epsilon}{2} (h^{\scaleto{\rm inh}{4pt}})^{ab} \partial_a \varphi \partial_b \varphi + O(\epsilon^2)~. 
\end{aligned}
\end{equation}
Note that we have discarded the homogeneous terms, as they contribute as total derivatives, i.e., they are pure gauge. The inhomogeneous solution can always be split into a trace ($\hat h$) and symmetric-traceless piece ($h^{\scaleto{\rm ST}{4pt}}_{ab}$),
\begin{equation}\label{eq:hab-h-ST}
    h^{\scaleto{\rm inh}{4pt}}_{ab}= \frac{1}{2}\bar g_{ab}\, \hat h + h^{\scaleto{\rm ST}{4pt}}_{ab}~.
\end{equation}
It is simple to check that the trace $\hat h$ always drops out from \eqref{eq:int-mass}, leaving us with 
\begin{equation}\label{eq:int-mass-1}
\begin{aligned}
    ({\cal L}_{\rm int})_{\mathfrak{m}^2=0} =&  - \frac{\epsilon}{2\Phi_0\ell_2^2}  \Y \varphi^2   + \frac{\epsilon}{2} (h^{\scaleto{\rm ST}{4pt}})^{ab} \partial_a \varphi \partial_b \varphi + O(\epsilon^2)~. 
\end{aligned}
\end{equation}
Next, the symmetric-traceless component of a two-dimensional tensor can always be cast locally as
    \begin{equation} \label{eq:hab-in-U}
        h_{ab}^{\rm ST} = \bn_a\bn_b U - \frac{1}{2}\bar{g}_{ab} \bar{\square} U~,
    \end{equation}
where $U(x)$ is a scalar function. With this, we note that 
    \begin{equation}
    \begin{aligned}
      \sqrt{-\bar g}\, (h^{\scaleto{\rm ST}{4pt}})^{ab}\partial_a \varphi \partial_b \varphi &= \sqrt{-\bar g}\,\bar{\nabla}^a \bar{\nabla}^b U(x) \left( \partial_a \varphi \partial_b \varphi - \frac{1}{2} \bar{g}_{ab} (\partial \varphi)^2\right) \\ 
        &= -\sqrt{-\bar g}\, \bar{\nabla}^b U(x) \left(\bar{\square} \varphi\right) \partial_b \varphi + ({\rm tot. ~der.})~,
    \end{aligned}
    \end{equation}
which is zero to leading order in $\epsilon$. Therefore, the only term that survives for massless fields is
\begin{equation}\label{eq:int-massless}
\begin{aligned}
    ({\cal L}_{\rm int})_{\mathfrak{m}^2=0} =&  - \frac{\epsilon}{2\Phi_0\ell_2^2}  \Y \varphi^2   + O(\epsilon^2)~. 
\end{aligned}
\end{equation}
It is rather unique that for massless fields, the effects of the near-AdS$_2$ background \eqref{eq:near-AdS2} rely only on how $\Y$ explicitly couples to the kinetic term of the scalar, and the metric has no imprint here. This indicates that massless fields have a universal backreaction that is model independent.\footnote{To this order in $\epsilon$, the details of the model of dilaton gravity enter via \eqref{eq:lin-dil}-\eqref{eq:ch-eq}, which is the place where details of $V(\Phi)$ have an imprint.} 

\paragraph{Massive fields ($\mathfrak{m}^2\neq 0$).} As we saw in \eqref{eq:int-scalar-gauge-2}, the essence of the backreaction effect on massive fields is due to the inhomogeneous terms in $h_{ab}$. This action can be simplified further by making the decomposition \eqref{eq:hab-h-ST}, which gives
\begin{equation}\label{eq:int-scalar-gauge-mass}
\begin{aligned}
    ({\cal L}_{\rm int})_{\mathfrak{m}^2\neq0} = &- \frac{\epsilon}{4} \mathfrak{m}^2 \,\hat h \, \varphi^2 + \frac{\epsilon}{2} (h^{\scaleto{\rm ST}{4pt}})^{ab} \partial_a \varphi \partial_b \varphi + O(\epsilon^2)\\
    = &- \frac{\epsilon}{4} \mathfrak{m}^2 \,\hat h \, \varphi^2 - \frac{\epsilon}{2} (h^{\scaleto{\rm ST}{4pt}})^{ab} \varphi (\bn_a  \bn_b \varphi) + \frac{\epsilon}{4} \varphi^2 \bn_a  \bn_b(h^{\scaleto{\rm ST}{4pt}})^{ab}    + O(\epsilon^2)
    ~. 
\end{aligned}
\end{equation}
Note that in the second line, we have ignored total derivatives. 

We now proceed to solve this part of \eqref{eq:lin-dil-2}. The simplest way to proceed is to pick coordinates of AdS$_2$ and make gauge choices to solve for  $h^{\scaleto{\rm inh}{4pt}}_{ab}$ explicitly. We will be interested in the near-AdS$_2$ backgrounds \eqref{eq:near-AdS2} that connect to black hole physics, which requires looking at static configurations. For this reason, we will describe AdS$_2$ as 
    \begin{equation} \label{eq:bh-bg}
        ds^2=\bar g_{ab} \dd x^a \dd x^b=\dd\rho^2+\frac{4\pi^2\ell_2^2}{\beta^2}\sinh^2(\rho/\ell_2) \dd\tau^2\,.
    \end{equation}
Note that here we have turned to Euclidean signature, and hence $\tau \sim \tau +\beta$ with $\beta$ an effective inverse temperature of AdS$_2$. In Lorentizian signature, we would have a horizon at $\rho=0$ and for this reason \eqref{eq:bh-bg} is known as the AdS$_2$ black hole. The dilaton $\Y$ in these situations is given by
    \begin{equation}\label{TimeIndepDilaton}
        \Y = \Y_0 \cosh(\rho/\ell_2)~,
    \end{equation}
which is the unique time-independent solution to \eqref{eq:Y-eom}.

To solve \eqref{eq:lin-dil-2}, we will impose $\hat h=0$, and the components of $h_{ab}^{\scaleto{\rm ST}{4pt}}$ to be\footnote{Indices are not summed over in this equation; it just indicates that each component of $h_{ab}^{\scaleto{\rm ST}{4pt}}$ is proportional to $\bar{g}_{ab}$.}
    \begin{equation}
        h_{ab}^{\scaleto{\rm ST}{4pt}}= \bar{g}_{ab}H_{ab}~,
        \label{eq:gauge-hst}
    \end{equation}
such that $h_{\tau\rho} = 0$ and the tracelessness of $h_{ab}^{\scaleto{\rm ST}{4pt}}$ imposes $ -H_{\tau \tau} =  H_{\rho\rho} =: H(x)$.  With these choices, \eqref{eq:lin-dil-2} reduces to $\bn^a\bn^b h_{ab}^{\scaleto{\rm ST}{4pt}} = \frac{\nu}{\ell_2^2} \Y $, which in components reads 
    \begin{equation} \label{eq:h-eq-components}
       \left(\frac{\beta^2}{4\pi^2\ell_2^2 \sinh^2(\rho/\ell_2)} \partial_\tau^2 - \partial_\rho^2 - \frac{3}{\ell_2}\coth(\rho/\ell_2) \partial_\rho - \frac{2}{\ell_2^2} \right) H(x) =- \frac{\nu}{\ell_2^2} \Y_0 \cosh(\rho/\ell_2)~.
    \end{equation}
The inhomogeneous solution is time-independent, with the resulting solution being
    \begin{equation}
        H(x)_{\scaleto{\rm inh}{4pt}} = \frac{\nu}{6}\Y_0 \cosh(\rho/\ell_2)~.
        \label{eq:sol-H}
    \end{equation}
We then find\footnote{As a side remark, we can check that indeed \eqref{eq:h-ST-explicit} is of the form \eqref{eq:hab-in-U}; the solution for $U(x)$ can be found explicitly, but is most succinctly written as
    \begin{equation}\nonumber 
        \bar{\square} U = \frac{2\nu}{3}\Y_0\cosh(\rho/\ell_2)\log\sinh(\rho/\ell_2)~. 
    \end{equation}}
    \begin{equation}\label{eq:h-ST-explicit}
        h_{ab}^{\scaleto{\rm ST}{4pt}} \dd x^a \dd x^b = \frac{\nu}{6}\Y_0 \cosh(\rho/\ell_2) \left( \dd \rho^2 - \frac{4\pi^2\ell_2^2}{\beta^2}  \sinh^2(\rho/\ell_2) \dd \tau^2 \right).
    \end{equation}
We can now evaluate \eqref{eq:int-scalar-gauge-mass}; using \eqref{eq:h-ST-explicit} gives
    \begin{equation}
        \begin{aligned}
         (h^{\scaleto{\rm ST}{4pt}})^{ab}  \varphi (\bn_a \bn_b\varphi) - \frac{1}{2} \varphi^2 \bn_a  \bn_b(h^{\scaleto{\rm ST}{4pt}})^{ab}  &=- \frac{\nu}{6}\Y\, \left(\varphi \bar\Box\varphi -2\varphi\partial_\rho^2 \varphi \right)-\frac{\nu}{2} \Y \varphi^2 +O(\epsilon)\\
           &=-\frac{\nu}{6} \Y\, \left(\mathfrak{m}^2\varphi^2-2\varphi\partial_\rho^2 \varphi \right)-\frac{\nu}{2}  \Y \varphi^2 +O(\epsilon)~,
        \end{aligned}
    \end{equation}
where in the last line we used \eqref{eq:gen-lin-scalar-eq}. As discussed before, the $\Y \varphi^2$ terms are pure gauge and should be discarded. Thus, in the black hole background, the interaction vertex for the massive fields becomes
    \begin{equation}\label{eq:massiveVertexGenForm}
        ({\cal L}_{\rm int})_{\mathfrak{m}^2\neq0} = -\epsilon\frac{\nu}{6}\Y\,\varphi\,\partial_\rho^2\varphi + O(\epsilon^2)\,.
    \end{equation}
The gauge choice leading to \eqref{eq:massiveVertexGenForm}, consisting of the choice for the background metric, dilaton, and independent components of $h_{ab}$, will be referred to as the ``black hole gauge''.  It is this expression that we will use in Sec.\,\ref{sec:corr-2pt-fn} to compute corrections to the two-point function of $\varphi$ due to the backreaction of AdS$_2$.

\section{Backreaction effects on two-point functions} \label{sec:corr-2pt-fn}

In this section, we will quantify how the backreaction induced by $\Y$ affects the correlation functions of the operator ${\cal O}_\Delta$, with $\Delta(\Delta-1)=\ell_2^2\mathfrak{m}^2$, dual to $\varphi$. As we saw in Sec.\,\ref{sec:general-2d-EA}, the effective action that describes this backreaction receives different simplifications depending on $\mathfrak{m}^2$. For a massless field, the action we will be working with is 
\begin{equation}\label{I-massless-EFT}
  (S_{\rm eff})_{\mathfrak{m}^2=0}=-\int \dd^2x\sqrt{\overbar{g}} \,\le(\frac{1}{2} \partial_a \varphi \partial^a \varphi  +\lambda_0\Y \varphi^2
 \ri),
\end{equation}
while for massive fields we have 
\begin{equation}\label{I-massive-EFT}
\begin{aligned}
     (S_{\rm eff})_{\mathfrak{m}^2\neq0}&=-\int \dd^2x\sqrt{\overbar{g}} \,\le(\frac{1}{2} \partial_a \varphi \partial^a \varphi + \frac{1}{2} \mathfrak{m}^2 \varphi^2 + \frac{\epsilon}{4} \mathfrak{m}^2 \,\hat h \, \varphi^2 - \frac{\epsilon}{2} (h^{\scaleto{\rm ST}{4pt}})^{ab} \partial_a \varphi \partial_b \varphi \ri) \\
     &=-\int \dd^2x\sqrt{\overbar{g}} \,\le(\frac{1}{2} \partial_a \varphi \partial^a \varphi + \frac{1}{2} \mathfrak{m}^2 \varphi^2 +\lambda_{\mathfrak{m}}\Y\varphi \partial_\rho^2\varphi
 \ri), 
\end{aligned}
    \end{equation}
where in the second line we wrote the interaction in the black hole gauge. The coupling constants $\lambda_0$ and $\lambda_\mathfrak{m}$ for the class of models \eqref{eq:gen-dil-model} are 
\begin{equation}
    \lambda_0=\frac{\epsilon}{2\Phi_0\ell_2^2}~,\qquad \lambda_{\mathfrak{m}} = \epsilon\frac{\nu}{6}= -\epsilon\frac{\ell_2^2}{6} V''(\Phi_0)~.
\end{equation}
In both massive and massless cases, we will treat $\Y$ as a time-independent background field in the black hole gauge \eqref{TimeIndepDilaton}, and evaluate the two-point function to leading order in $\epsilon$. The answer, in Euclidean signature, will be of the form
\begin{equation}\label{eq:2pt-func}
      \langle {\cal O}_\Delta(\tau){\cal O}_{\Delta}(0)\rangle =   \langle {\cal O}_\Delta(\tau){\cal O}_{\Delta}(0)\rangle_{\rm free} +  \langle {\cal O}_\Delta(\tau){\cal O}_{\Delta}(0)\rangle_{\epsilon} + O(\epsilon^2)
\end{equation}
where the free piece, of order $O(\epsilon^0)$, is 
\begin{equation}\label{eq:2pt-free}
    \langle {\cal O}_\Delta(\tau){\cal O}_{\Delta}(0)\rangle_{\rm free} =  \frac{(2\Delta-1)\Gamma(\Delta)}{\sqrt{\pi}\,\ell_2^2\,\Gamma(\Delta-\frac{1}{2})}\left(\frac{\pi\ell_2}{\beta} \frac{1}{\sin(\frac{\pi\tau}{\beta})}\right)^{2\Delta}  \,.
\end{equation}
As we will review, this free piece is the result of evaluating a correlation on thermal AdS$_2$, where the Euclidean time has period $\beta$. The main focus of this section is to evaluate $\langle {\cal O}_\Delta(\tau){\cal O}_{\Delta}(0)\rangle_{\epsilon}$, the leading order classical effect due to the backreaction of near-AdS$_2$.

A method to evaluate \eqref{eq:2pt-func} was first advocated in \cite{MalSta16} by treating the background value of $\Y$ as an operator with $\Delta_\Y=-1$ with a static source, and considering a simple cubic interaction of the form $\Y \varphi^2$. 
However, there is a problem with the analysis in \cite{MalSta16}.  
First, as we have seen in Sec.\,\ref{sec:gen-corr-2pt}, in more general dilaton theories, where one needs to account for the backreaction of the metric, the vertex $\Y\varphi^2$ is also accompanied by a coupling between the graviton and the scalar fields. Even when treated as a toy model, a cubic coupling of the form $\Y\varphi^2$ will not correct the two-point function: as we showed in \eqref{eq:int-scalar-Y-1}, such a vertex is pure gauge for massive fields. This is further corroborated in App.\,\ref{App:Vertex}, where we show explicitly that for massive fields ($\Delta > 1$) the contribution of $\Y \varphi^2$ to  $\langle {\cal O}_\Delta(\tau){\cal O}_{\Delta}(0)\rangle_{\epsilon}$ is zero when $\Y$ is treated as a dilaton with background value \eqref{TimeIndepDilaton}.

For massless fields, where $\Delta=1$, the answer reported in \cite{MalSta16} for the correction \eqref{eq:2pt-func} turns out to be correct; in this case, as can be seen from \eqref{eq:int-massless}, the vertex is indeed of the form $\Y\varphi^2$. 
However, there is \textit{prima facie} still a problem with the proposed method: if we treat $\Y$ as an operator with $\Delta_\Y=-1$ in \eqref{I-massless-EFT}, the interaction is then extremal. As noted in \cite{CasVer21,CasMar24}, this leads to a divergence in the cubic interaction computed in this way, and again the method seems to break down. In App.\,\ref{App:Vertex} we comment on why the result reported in \cite{MalSta16} works for massless fields: carefully keeping track of the normalisation constants cancels the divergence.

\subsection{Preliminaries}

Here we collect some useful formulae and fix some notation, before moving on to the explicit evaluation of \eqref{eq:2pt-func} for massless and massive fields on the black hole gauge.  
To perform the computation carefully, we work in Euclidean signature on a regulated AdS$_2$ black hole space foliated as \eqref{eq:bh-bg}
\begin{equation}
 ds^2=\dd\rho^2+\frac{4\pi^2\ell_2^2}{\beta^2}\sinh(\rho/\ell_2)^2 \dd\tau^2 ~,\qquad\rho\in[0,\rho_c]~, \qquad\tau\sim\tau+\beta\,.
\end{equation}
where $\rho_c$ is an IR cut-off and $\beta$ is the inverse temperature of the AdS$_2$ black hole. The induced metric at $\rho=\rho_c$, the normal derivative and induced Laplacian at the boundary are
\begin{equation}\label{gamma-dn} 
\sqrt{\gamma}=\frac{2\pi\ell_2}{\beta} \sinh(\rho_c/\ell_2)~,\qquad \partial_n=+\partial_\rho~,
\qq\square_\gamma=\frac{\beta^2}{4\pi^2\ell_2^2}\frac{\partial_\tau^2 }{\sinh(\rho_c/\ell_2)^2} ~.
\end{equation}
It is also useful to define
\begin{equation}
    \Lambda:= \frac{2\pi \ell_2}{\beta}\,\sinh(\rho_c/\ell_2)~,
\end{equation}
which is the Weyl factor defining the boundary metric, i.e.,
\begin{equation}
d\tau^2=\lim_{\rho_c\to\infty} \Lambda^{-2} ds^2|_{\rho=\rho_c}~.
\end{equation}
The dilaton in this gauge reads
\begin{equation}\label{TimeIndepDilaton-repeat}
{\cal Y}(\rho,\tau)= {\cal Y}_0 \cosh(\rho/\ell_2)~, 
\end{equation}
where ${\cal Y}_0$ is a constant that fixes the dilaton at the horizon, i.e., ${\cal Y}(0,\tau) = {\cal Y}_0$. This amounts to a renormalised boundary value $\Y_b=\frac{\beta}{2\pi\ell_2}\Y_0$ for the dilaton in this gauge.

In the absence of an interaction, that is, setting $\epsilon=0$ in $S_{\rm eff}$, it is simple to write the scalar field in terms of the bulk-to-boundary propagator. By imposing  Dirichlet boundary conditions at $\rho=\rho_c$ in the black hole gauge, the field reads
\begin{equation}\label{Scalar-BH-sol}
     \varphi(\rho,\tau)_{\rm free}=\sum_{n\in\mathbb{Z}} K_{\Delta;|n|}(\rho) e^{-i \frac{2\pi n}{\beta} \tau }\tilde\varphi_{n}~, \qq \varphi(\rho_c,\tau)_{\rm free}=\Lambda^{\Delta-1}\tilde\varphi(\tau)~,
\end{equation}
where
\begin{align}\label{Scalar-General-K}
K_{\Delta;n}(\rho)
 =\Lambda^{\Delta-1}
    \frac{P_{\Delta -1}^{-|n|}\left(\cosh \left(\frac{\rho}{\ell_2} \right)\right)}{P_{\Delta -1}^{-|n|}\left(\cosh \left(\frac{\rho_c}{\ell_2} \right)\right)}~,
    \end{align}
and $P_{n}^{m}(x)$ is the associated Legendre function. Note that throughout this section, we will take $\Delta$ to be the largest root of $\Delta(\Delta-1)=\mathfrak{m}^2$. Here $\tilde\varphi(\tau)$ is the source for the field, and its Fourier transform is defined as
\begin{equation}\label{FTransform} 
\tilde\varphi_{n}\equiv \int_0^{\beta}\!\!\frac{\dd \tau}{\beta}\, e^{i \frac{2\pi n}{\beta} \tau }\,\tilde\varphi(\tau)~,
\qquad \tilde\varphi(\tau)\equiv \sum_{n\in\mathbb{Z}} e^{-i \frac{2\pi n}{\beta} \tau }\tilde\varphi_{n}~.
\end{equation}
With this set of functions, we can obtain the first-order correction in $\epsilon$ to the two-point functions due to the backreaction of AdS$_2$ following the standard AdS/CFT procedures; see, e.g., \cite{Mue99,BotMar17}. In short, the principle is to replace \eqref{Scalar-BH-sol} in the effective actions \eqref{I-massless-EFT} and \eqref{I-massive-EFT} respectively, evaluate the appropriate bulk integrals, and then read off the two-point function as a variation with respect to  $\tilde\varphi(\tau)$. In the following subsections, we will carry out this procedure in detail for both massless and massive fields separately. 

\subsection{Two-point function massless scalar (\texorpdfstring{$\Delta=1$}{Delta = 1})} \label{sec:massless-explicit-vertex}

We begin by evaluating \eqref{eq:2pt-func} for a massless field, where we adopt the quantisation condition such that $\Delta=1$. In this case,  the bulk-to-boundary propagator \eqref{Scalar-General-K} reduces to
\begin{equation}\label{eq:K-massless}
    K_{1;|n|}(\rho)=\left(\frac{\tanh \left(\frac{\rho }{2 \ell_2}\right)}{ \tanh
\left(\frac{\rho_c}{2 \ell_2}\right)}\right)^{|n|}~.
\end{equation}
Next, treating the field as an expansion in $\epsilon$, we write
\begin{equation}
    \varphi(x) = \varphi(x)_{\rm free} + O(\epsilon)~,
\end{equation}
and, to leading order in $\epsilon$, evaluating \eqref{I-massless-EFT} on-shell gives 
\begin{equation}\label{I0-massless-EFT}
  (I_{\rm eff})_{\mathfrak{m}^2=0}= -\frac 12\int \dd\tau \sqrt{ \gamma} \,\vo \partial_n \vo -\lambda_0 \int \dd^2x\sqrt{\overbar{g}} \;\Y \vo^2 + O(\epsilon^2) \,.
\end{equation}
The first term here is the free contribution, and reading off the two-point function is standard. Using \eqref{gamma-dn}, \eqref{Scalar-BH-sol} and \eqref{eq:K-massless} gives
\begin{equation}\label{eq:free-massless-2pt}
\begin{aligned}
-\frac 12 \int\!\! \dd\tau\sqrt{\gamma}\, \vo \partial_n \vo&= -\frac{1}{2}\sum_{n\in\mathbb{Z}}\vonnt(\beta \sqrt{\gamma}\, K'_{1;|n|}(\rho))_{\rho=\rho_c}\\
&= -\frac{1}{2}\sum_{n\in\mathbb{Z}}\vonnt \, (2 \pi |n|)\\
&
=\frac 12 \int_0^\beta  \!\!\! \dd\tau\int_0^\beta\!\!\!\dd\tau'\; \vot(\tau) \left(\frac{\pi  }{\beta ^2}\frac{1}{\sin ^2\left(\frac{\pi 
   (\tau-\tau')}{\beta }\right)}\right)\vot(\tau')\;.
\end{aligned}
\end{equation}
We get the standard free correlator with its standard holographic normalisation, i.e.,
\begin{equation}\label{2pf-O0}
   \langle {\cal O}_1(n){\cal O}_1(-n)\rangle_{\rm free} = 2\pi |n| \;,\qq \langle {\cal O}_1(\tau){\cal O}_1(0)\rangle_{\rm free}
    =\frac 1\pi\left(\frac{\pi }{\beta}\frac{1}{\sin\left(\frac{\pi 
   \tau}{\beta }\right)}\right)^2~,
\end{equation}
in agreement with \eqref{eq:2pt-free}.
We now consider the contribution from the interaction term in \eqref{I0-massless-EFT}. Using \eqref{TimeIndepDilaton-repeat}, \eqref{Scalar-BH-sol}, and \eqref{eq:K-massless} gives to leading order in $\epsilon$ 
\begin{equation}
    \begin{aligned}
        \label{VertexInt-1}
-\lambda_0 \int \dd^2 x  \sqrt{\overline g} \, {\cal Y}\,\vo^2
&=-2\pi\ell_2^2\lambda_0\Y_0  \sum_{n\in\mathbb{Z}}\vonnt\left( \frac{e^{2\frac{\rho_c}{\ell_2}}}{8}-e^{\frac{\rho_c}{\ell_2}}\frac{|n|}{2} -\frac 14 +n^2 \log \left(\frac{e^{2\frac{\rho_c}{\ell_2}}}{16}\right) \right. \\
&\qq\qq\qq\qq\qq\qq\qq\qq\left. +|n|\left(1 -2|n| H_{|n|}\right)  +O\left(e^{-\frac{\rho_c}{\ell_2}}\right)\right)
    \end{aligned}
\end{equation}
where $H_n$ are harmonic numbers.
The first line in \eqref{VertexInt-1} involves counterterms, and the second line contains the correction to the correlator.  In momentum space, the correction reads
\begin{equation}\label{2pf-0momentum}
    \langle {\cal O}_1(n){\cal O}_1(-n)\rangle_{\epsilon}
    =4\pi\ell_2^2\lambda_0 \mathcal{Y}_0 |n| \left(1-2|n|H_{|n|}\right).
\end{equation}
It is instructive to transcribe this to configuration space. The term proportional to $|n|$ is just a renormalisation of the correlator, as seen in \eqref{eq:free-massless-2pt}. 
The second contribution in \eqref{2pf-0momentum} Fourier transforms to 
\begin{equation}
    \begin{aligned}\label{CorrJT2}
     &\sum_{n}\vonnt (2 n^2 H_{|n|})\\
    &\qquad=\frac{1}{2\pi} \int\dd \tau\dd \tau'\,\vot(\tau) \left(\frac{\pi  }{\beta }\frac{1}{\sin\left(\frac{\pi 
   }{\beta }(\tau-\tau')\right)}\right)^2 \left(1+
   \pi\frac{ 1-2|\tau-\tau'|/\beta }{\tan
   \left(\frac{\pi 
   }{\beta }|\tau-\tau'|\right)}\right) \vot(\tau')
    \end{aligned}
\end{equation}
Using \eqref{eq:free-massless-2pt} and \eqref{CorrJT2} we get that the two-point function in configuration space is
\begin{equation}\label{on-shell-lambda}
\begin{aligned}
        (I_{\rm eff})_{\mathfrak{m}^2=0}=\frac12\int \dd \tau \dd \tau' \, \vot(\tau) \, \langle\mathcal{O}_1(\tau)\mathcal{O}_1(\tau')\rangle\,\vot(\tau')
\end{aligned}
\end{equation}
where
\begin{equation}
\begin{aligned}\label{Delta1corrected}
\frac{\langle\mathcal{O}_1 (\tau) \mathcal{O}_1(0)\rangle}{\langle\mathcal{O}_1 (\tau) \mathcal{O}_1(0)\rangle_{\rm free}}&=1+2\ell_2^2 \lambda_0 {\cal Y}_0\left(2+\pi\frac{1-2\tau/\beta}{\tan(\frac{\pi \tau}{\beta})}\right)+ O(\epsilon^2)\\
&=1+ \frac{\epsilon{\cal Y}_0}{\Phi_0}\left(2+\pi\frac{1-2\tau/\beta}{\tan(\frac{\pi \tau}{\beta})}\right)+ O(\epsilon^2)~,
\end{aligned}
\end{equation}
where above $\tau\in[0,\beta)$ and the function is defined as periodic $\tau\sim\tau+\beta$ outside this domain. In the second line, we replaced the value of $\lambda_0$ for the appropriate value arising from the effective dilaton model of Sec.\,\ref{sec:general-2d-EA}; note that the correction is suppressed by the extremal classical entropy ($\Phi_0$) and the strength of the backreaction ($\epsilon {\cal Y}_0$). 
The result in \eqref{Delta1corrected} is in agreement with the functional dependence in \cite{MalSta16} when written in terms of $\Y_b$, given below \eqref{TimeIndepDilaton-repeat}.
Still, it is important to stress that na\"ively, using the bulk-to-boundary propagator \eqref{Scalar-General-K} with $\Delta_\Y = -1$ rather than \eqref{TimeIndepDilaton-repeat} yields a divergent answer due to the correlator being extremal as reported in \cite{CasVer21}; only a careful assessment of the normalisation constants will lead to a cancellation of the divergences and a finite answer, see App.~\ref{App:Vertex}.

We conclude this part by coming back to the logarithmic term in \eqref{VertexInt-1}. This term indicates that the interaction between the dilaton and the massless field induces an anomaly of the form, see e.g. \cite{deHaro:2000vlm,Papadimitriou:2016yit,MalSta16}
\begin{equation}\label{JT-Anomaly}
(S_{\rm eff})_{\mathfrak{m}^2=0}\supset \frac{\rho_c}{2}\int \dd\tau {\cal A}~, \qq\qq {\cal A}
=4\ell_2^3\lambda_0\Y_b\vot(\tau)\vot''(\tau)\;.
\end{equation}
One can put this alongside the standard Schwarzian contribution to the stress tensor, resulting in a Callan--Symanzik equation for this theory that reads\footnote{There should also be a quantum contribution to $\langle T(\tau)\rangle_s $ coming from the path integral of the Schwarzian action. Including quantum effects is more delicate and will not be discussed here. }
\begin{equation}
\begin{aligned}\label{CalSym-JT-beta}
    \langle T(\tau)\rangle_s 
    &=\epsilon\frac{ \Y_b}{\kappa_2^2}\ell_2\;
    \text{Sch}(\tan( \pi \tau/\beta),\tau)+4\ell_2^3\lambda_0\Y_b \vot(\tau)\vot''(\tau)+ \ldots\\
    &=  
    \epsilon\frac{ \Y_b}{\kappa_2^2}\; \frac{2 \pi ^2\ell_2}{\beta ^2}+ 2\epsilon \ell_2\frac{{\cal Y}_b}{\Phi_0}\,\vot(\tau)\vot''(\tau) +\ldots~,
\end{aligned}
\end{equation}
The anomaly in \eqref{JT-Anomaly} is independent of $\beta$ and suppressed by a factor of $\Phi_0$ with respect to the Schwarzian contribution. Notice that for constant sources the anomaly vanishes. Anomalies that vanish whenever matter sources are turned off are often referred to as ``matter conformal anomalies'' and are common in the AdS/CFT literature, see, e.g., \cite{deHaro:2000vlm,Ske02}.

\subsection{Two-point function massive scalars (\texorpdfstring{$\Delta>1$}{Delta >1})} \label{sec:massive-explicit-vertex}

The interaction between the massive field and the dilaton is qualitatively different from the massless case and should be treated separately. Our starting point is \eqref{I-massive-EFT}: working in the black hole gauge and using \eqref{Scalar-BH-sol}-\eqref{Scalar-General-K}, we find that the on-shell action to leading order in $\epsilon$ is
\begin{equation}\label{I0-massive-EFT}
         (I_{\rm eff})_{\mathfrak{m}^2\neq0}=-\frac 12\int \dd\tau\sqrt{ \gamma} \,\vD \partial_n \vD -\lambda_{\mathfrak{m}}\int \dd^2x\sqrt{\overbar{g}}\;\Y \,\vD \partial_\rho^2\vD+ O(\epsilon^2)\,. 
    \end{equation}
Evaluating this cubic term for generic values of $\mathfrak{m}$, or $\Delta$, is difficult. To illustrate the outcome of the interaction in the two-point function \eqref{eq:2pt-func}, we will start by taking explicit integer values of $\Delta$, and then generalise the answer for any $\Delta\in \mathbb{Z}^+$.  In subsequent sections, by using the comparison to correlators on BTZ, we will comment on the results for any real $\Delta\geq0$.

To start, let us take the simplest integer value for the massive field: $\Delta=2$, $\ell_2^2\mathfrak{m}^2=2$. The propagator \eqref{Scalar-General-K} simplifies in this case to
\begin{equation}
   K_{2;|n|}(\rho)=\Lambda 
   \frac{\left(|n|+\cosh \left(\frac{\rho}{\ell_2}\right)\right) \tanh ^{|n|}\left(\frac{\rho}{2\ell_2}\right)}{\left(|n|+\cosh \left(\frac{\rho_c}{\ell_2}\right)\right) \tanh ^{|n|}\left(\frac{\rho_c}{2\ell_2}\right)}\;. 
\end{equation}
With this solution at hand, we can check that the free correlator yields
\begin{equation}\label{FreeCorr-2}
\begin{aligned}
-\frac 12 \int\!\! \dd\tau \sqrt{\gamma}\, \vd \partial_{\rho} \vd &=-\frac{1}{2}\sum_{n\in\mathbb{Z}}\vdnnt (\beta\sqrt{\gamma}\Lambda K_{2;|n|}'(\rho))_{\rho=\rho_c}\\
   &=\frac 12 \int \dd\tau\dd\tau'\vdt(\tau)\left(\frac 6\pi \,\ell_2^2\! \left(\frac{\pi}{\beta \sin\left(\frac{\pi(\tau-\tau')}{\beta}\right)}\right)^{4}\right)\vdt(\tau')-I_{\rm ct}^{(0)}~,
\end{aligned}
\end{equation}
with $I_{\rm ct}^{(0)}$ a standard set of counterterms. Therefore, the two-point function is
\begin{equation}
    \langle {\cal O}_2(\tau){\cal O}_2 (0)\rangle_{\rm free} =\frac 6\pi \,\ell_2^2\! \left(\frac{\pi }{\beta}\frac{1}{\sin \left(\frac{\pi 
   \tau}{\beta }\right)}\right)^4~,
\end{equation}
which matches with \eqref{eq:2pt-free}. For the interaction term in \eqref{I0-massive-EFT} we get
\begin{equation}\label{VertexInt-2}
\begin{aligned}
   -\lambda_\mathfrak{m}\int \dd^2x\sqrt{\bar g}\,  \Y\vd\partial_\rho^2\vd 
&=-\lambda_\mathfrak{m}{\cal Y}_0\int \dd^2 x  \sqrt{\overline g}
\,\cosh(\rho/\ell_2) \vd (\partial^2_\rho\vd )\\
&=\frac{4\pi^3 \ell_2^2 }{\beta ^2} \lambda_\mathfrak{m}  {\cal Y}_0 \sum_{n\in\mathbb{Z}}\vdnnt\; |n| \left(n^2-1\right) \left(1-2 |n| H_{|n|}\right)-I_{\rm ct}^{(1)}~, 
\end{aligned}
\end{equation}
where the counterterms $I_{\rm ct}^{(1)}$ include once again a log term.
This term induces an anomaly that can be straightforwardly computed as we did in the $\Delta=1$ case in \eqref{JT-Anomaly}-\eqref{CalSym-JT-beta}. These are again matter conformal anomalies and are present for all integer $\Delta\geq1$. We can Fourier transform the first term in the second line of \eqref{VertexInt-2} to configuration space to obtain a corrected two-point function. This gives
\begin{equation}\label{Delta2corrected}
    \frac{\langle {\cal O}_2(\tau){\cal O}_2(0)\rangle}{\langle {\cal O}_2(\tau){\cal O}_2 (0)\rangle_{\rm free}} =   1+ \frac{2}{3} \lambda_\mathfrak{m}  {\cal Y}_0  \left(5+3 \pi  \frac{(1 -2 \tau/\beta)}{\tan \left(\frac{\pi  \tau }{\beta }\right)} +\cos \left(\frac{2 \pi  \tau }{\beta }\right) \right)  + O(\epsilon^2)~.
\end{equation}
The functional dependence on $\tau$ here is different from the one reported in \cite{MalSta16}.

For any integer $\Delta\geq2$, evaluating $\langle {\cal O}_\Delta(\tau){\cal O}_\Delta(0)\rangle$ to leading order in $\epsilon$ follows similarly to the above derivation. An interesting feature is that, in momentum space, a simple pattern unifies the corrections to the two-point function. We can write the effective action for integer $\Delta$ as 
\begin{equation}\label{Gn-Def}
    (I_{\rm eff})_{\mathfrak{m}^2}=-\frac12 \sum_{n\in\mathbb{Z}}\vdnnt \Big( \langle {\cal O}_\Delta (n){\cal O}_\Delta(-n)\rangle_{\rm free} +c_\Delta \langle {\cal O}_\Delta (n){\cal O}_\Delta(-n)\rangle_{\epsilon}\Big)~, 
\end{equation}
where we have unified the massive and massless cases via
\begin{equation}\label{def:c-delta}
c_\Delta=\begin{cases}2\ell_2^2 \lambda_0{\cal Y}_0~,\quad    & \Delta=1~,\\ \lambda_{\mathfrak{m}}{\cal Y}_0~,\quad  & \Delta=2,3,\dots~.        
    \end{cases} 
\end{equation}
The free contribution is the standard expression
\begin{equation}
\begin{aligned}
   \langle {\cal O}_\Delta (n){\cal O}_\Delta(-n)\rangle_{\rm free} 
     &=\left(\frac{\pi\ell_2}{\beta}\right)^{2\Delta-2}\frac{2\pi^2 |n|}{\Gamma\left(\Delta-\frac{1}{2}\right)^2}\prod_{m=1}^{\Delta-1}(m^2-n^2)\,.
    \label{free piece}
\end{aligned}
\end{equation} 
One finds that the correction to the two-point function due to the backreaction of AdS$_2$ has the simple form 
\begin{equation}\label{relationGcorrGfree}
\langle {\cal O}_\Delta (n){\cal O}_\Delta(-n)\rangle_{\epsilon} 
    =  (1-2|n|H_{|n|})\langle {\cal O}_\Delta (n){\cal O}_\Delta(-n)\rangle_{\rm free} \;.
\end{equation}
To obtain the correction to the two-point function in configuration space for $\Delta\in \mathbb{Z}^+$,  we can exploit the simple structure \eqref{relationGcorrGfree} alongside the standard recursion for the free correlator 
\begin{equation}
     \langle {\cal O}_{\Delta+1} (n){\cal O}_{\Delta+1}(-n)\rangle_{\rm free} 
   =\left(\frac{ \pi  \ell_2 }{\beta}\right)^2\frac{\left(\Delta^2
  -n^2 \right)}{\left( \Delta-\frac{1}{2} \right)^2}  \langle {\cal O}_\Delta (n){\cal O}_\Delta(-n)\rangle_{\rm free} ~.
\end{equation}
It turns out that this recursion relation can be explicitly solved; the details can be found in App.\,\ref{app:rec-rel}. The final result is
\begin{equation} \label{eq:corr-explicit}
  \begin{aligned}
   \langle {\cal O}_\Delta (\tau){\cal O}_\Delta(0)\rangle_{\epsilon}&= \frac{1}{\beta^2}\sum_{n\in\mathbb{Z}}e^{i \frac{2\pi n}{\beta}\tau}\langle {\cal O}_\Delta (n){\cal O}_\Delta(-n)\rangle_{\epsilon}  \\
  & =
   \,\langle {\cal O}_\Delta (\tau){\cal O}_\Delta(0)\rangle_{\rm free}\left(2 + \Delta \pi \frac{1-2\tau/\beta}{\tan(\frac{\pi \tau}{\beta})}   +   S_{\Delta} (\tau) \right)~,
  \end{aligned}
\end{equation}
where
    \begin{equation} \label{eq:S-Delta}
   \begin{aligned}
        S_{\Delta} (\tau) =&~ \frac{2\sqrt{\pi}\,\Gamma(\Delta)}{(2\Delta-1)\Gamma(\Delta-\frac{1}{2})}\sum_{m=1}^{\Delta-1} 2m(2m+1)(-1)^m \,e_m\Big(1, \frac{1}{2}, \ldots, \frac{1}{(\Delta-1)^2}\Big) \times \\
        &\times \sum_{k=0}^{2m} \sum_{j=0}^{k-1} \frac{(-1)^j(k-j)^{2m}}{(k+1)2^{2k}\sin(\frac{\pi\tau}{\beta})^{2k+2-2\Delta}}\binom{2m}{k}\binom{2k}{j} 
        \cos\left(2(k-j)\frac{\pi\tau}{\beta}\right)~.
    \end{aligned}        
    \end{equation}
Here $e_m(\ldots)$ denotes the elementary symmetric polynomial, see \eqref{eq:sym-pol-def} for a definition. Hence, for $\Delta\in \mathbb{Z}^+$ we have
\begin{equation}\label{Delta-integer-corrected}
    \frac{\langle {\cal O}_\Delta(\tau){\cal O}_\Delta(0)\rangle}{\langle {\cal O}_\Delta(\tau){\cal O}_\Delta (0)\rangle_{\rm free}} =   1+ c_\Delta   \left(2 + \Delta  \pi \frac{1-2\tau/\beta}{\tan(\frac{\pi \tau}{\beta})}   +   S_{\Delta} (\tau) \right)  + O(\epsilon^2)~,
\end{equation}
with $c_\Delta$ given in \eqref{def:c-delta}. This expression reduces correctly to the massless \eqref{Delta1corrected} and massive ($\Delta = 2$) \eqref{Delta2corrected} cases. 

As mentioned around \eqref{Delta1corrected}, we stress that the result \eqref{eq:corr-explicit} agrees only for $\Delta = 1$ with \cite{MalSta16}, for which $S_{\Delta}(\tau) = 0$. According to \cite{MalSta16}, the correction coming from a ${\cal Y}\varphi^2$ bulk vertex for general $\Delta$ is 
    \begin{equation}
          \frac{\langle {\cal O}_\Delta(\tau){\cal O}_\Delta(0)\rangle}{\langle {\cal O}_\Delta(\tau){\cal O}_\Delta (0)\rangle_{\rm free}} = 1+ a_0\, \lambda\,\Y_0\,   \left(2 +\pi\frac{1-2\tau/\beta}{\tan(\frac{\pi \tau}{\beta})}\right) + O(\lambda^2)\,,
    \end{equation}
where $\lambda$ is their non-derivative coupling between $\Y$ and the scalar fields, and $a_0$ is a numerical constant. We see that the functional form of their correction is similar to the first two terms of \eqref{eq:corr-explicit}; however, by evaluating the vertex \eqref{eq:massiveVertexGenForm}, which captures correctly the near-extremal backreaction for a massive field, we find a $\Delta$-dependent answer, and furthermore, the non-trivial term $S_{\Delta}(\tau)$ appears. As mentioned previously, evaluating the vertex $\Y\varphi^2$ for massive fields will yield zero, since this vertex is pure gauge. We explicitly show this in App.~\ref{App:Vertex}. 

Evaluating \eqref{VertexInt-2} for real (non-integer) $\Delta$ is not simple.  This is because the scalar propagator \eqref{Scalar-General-K} reduces to polynomials for integer $\Delta$, but in general they are complicated series expansions. In the coming sections, we will be able to infer the generalisation of \eqref{Delta-integer-corrected} by using the comparison to the two-point correlation function on BTZ. 

\section{The backreaction of near-extremal BTZ} \label{sec:BTZ-BHs}

In this section, we will test our findings in Sec.\,\ref{sec:general-2d-EA} and Sec.\,\ref{sec:corr-2pt-fn} against well-known expressions of the two-point function of a scalar operator on the BTZ black hole. The setup is simple: we will be working with the Einstein--Hilbert action coupled to a massive scalar field with action  
\begin{equation}\label{eq:action-3d}
    S_{\rm 3D}= \frac{1}{2\kappa_3^2}\int \dd^3 x \sqrt{-G}\,\left( R^{(3)} + \frac{2}{\ell_3^2} - \frac{1}{2}\partial_\mu \varphi \partial^\mu \varphi - \frac{\mathfrak{m}^2}{2} \varphi^2 \right).
\end{equation}
Here $\kappa_3^2=8\pi G_3$ and $\ell_3$ is the AdS$_3$ radius. The scalar field is dual to an operator ${\cal O}_{\rm 3D}$, with conformal dimension $\Delta_{\rm 3D}(\Delta_{\rm 3D}-2)= \ell_3^2 \mathfrak{m}^2$. To leading order in $G_3$, we will be inspecting the retarded two-point function of ${\cal O}_{\rm 3D}$ on BTZ.  In particular, we will take a low-frequency and low-temperature limit of this two-point function and contrast the leading and subleading corrections to our findings in Sec.\,\ref{sec:corr-2pt-fn}. In short, we will show that the subleading correction at low-frequency/low-temperature is exactly the effect of the backreacted geometry of the near-extremal near-horizon geometry of BTZ.

We will start this section by first reviewing the $s$-wave sector of the Kaluza--Klein reduction of the action \eqref{eq:action-3d}. In particular, we will relate it to the dilaton model in \eqref{eq:gen-dil-model} and translate the details of the near-extremal BTZ black hole into this two-dimensional notation. With this information, we will report on the precise prediction that the analysis of Sec.\,\ref{sec:corr-2pt-fn} makes on the two-point function in the near-AdS$_2$ region of BTZ.  Then we will manipulate the retarded two-point function of ${\cal O}_{\rm 3D}$ obtained from the perspective of AdS$_3$/CFT$_2$, and find perfect agreement between these expressions in the near-extremal limit. 

\subsection{Near-extremal BTZ from a two-dimensional perspective}

Several references discuss three-dimensional gravity from a two-dimensional perspective; see, for example, \cite{AchOrt93,CvePap16,AlmKan16,GhoMax19}. Here, we will provide a brief account to establish the notation relative to Sec.\,\ref{sec:general-2d-EA}.   
To perform the Kaluza--Klein reduction, we take the metric ansatz
\begin{equation}
    ds^2_3 = G_{\mu \nu}\dd x^\mu \dd x^\nu =g_{ab} \dd x^a \dd x^b + \Phi^2 (\dd\phi + A_a \dd x^a)^2~,
    \label{metric ansatz}
\end{equation}
where $\phi\sim\phi+2\pi$ is the direction of compactification. The fields $g_{ab}$, $\Phi$, and $A_a$ will depend only on the two-dimensional directions $x^a$, $a=1,2$. We will also assume that the massive field $\varphi$ depends only on $x^a$.  Then, the dimensional reduction of \eqref{eq:action-3d} yields the two-dimensional action
    \begin{equation}
  S_{\rm 2D} = \frac{\pi}{\kappa_3^2} \int \dd^2 x \sqrt{-g} \,\Phi \left(R^{(2)} + \frac{2}{\ell_3^2} - \frac{1}{4} \Phi^2 F_{ab}F^{ab} - \frac{1}{2} g^{\mu \nu} \partial_\mu \varphi \partial_\nu \varphi - \frac{1}{2}\mathfrak{m}^2 \varphi^2\right).
  \label{eq:2D-3Daction}
\end{equation}
where  $F=\dd A$. Note that contrary to the discussion in Sec.\,\ref{sec:general-2d-EA}, here $\Phi$ has units of length, as is the case in many dimensional reductions: this can be seen directly from \eqref{metric ansatz}. Thus, $\Phi/\kappa_3^2$ is dimensionless and it is related to  $\Phi/\kappa_2^2$ in Sec.\,\ref{sec:general-2d-EA}.

Next, we can easily integrate out the gauge field because we are in the $s$-wave sector, for which the scalar field $\varphi$ does not couple to $A_a$. Explicitly, the gauge field in this sector is simply 
\begin{equation}
    F_{ab}= \frac{Q}{\Phi^{3}} \,\epsilon_{ab}~, 
\end{equation}
with $\epsilon_{ab}$ the Levi-Civita tensor, and $Q$ a constant.
The resulting action, after doing an appropriate Legendre transformation to integrate out $A_a$, is  
\begin{equation}
  S_{\rm 2D} = \frac{\pi}{\kappa_3^2} \int \dd^2 x \sqrt{-g} \,\Phi \left(R^{(2)} + \frac{2}{\ell_3^2} - \frac{Q^2}{2\Phi^4} - \frac{1}{2} g^{\mu \nu} \partial_\mu \varphi \partial_\nu \varphi - \frac{1}{2}\mathfrak{m}^2 \varphi^2\right).
  \label{reduced action}
\end{equation}
We can now easily see that \eqref{reduced action} is of the form \eqref{eq:gen-dil-model}, with a potential given by
\begin{equation}
   V(\Phi) = \frac{2\Phi }{\ell_3^2} - \frac{Q^2}{2\Phi^3}~.
\end{equation}
Recall that the AdS$_2$ vacuum required that $V(\Phi_0) = 0$ and $V'(\Phi_0)$ defines the AdS$_2$ radius via \eqref{eq:eom-ads2}. These two conditions gives 
\begin{equation}\label{eq:Phi0-3d}
\Phi_0^2 = |Q|\ell_2  ~, \qquad \ell_3 = 2 \ell_2~.  
\end{equation} 
 From here, we can also read the coupling term appearing in the backreaction of the metric. From \eqref{eq:ch-eq}, we have
 \begin{equation}
 \nu = -\frac{6}{\Phi_0}~.    
 \end{equation}

\paragraph{Near-Extremal BTZ.} We will now recast the BTZ black hole \cite{Banados:1992wn}  in the 2D language, and relate the backreaction of AdS$_2$ to the near-extremal geometry of BTZ. We will start by reviewing some basic properties of the black hole. The metric of non-extremal BTZ is given by
\begin{equation}
    ds^2 = - f(r) \dd t^2 + \frac{\dd r^2}{f(r)} + r^2 \left(\dd\psi + N^\psi(r) \dd t\right)^2~. 
\end{equation}
Here the blackening factor $f(r)$ and angular dragging $N^\psi(r)$ are 
\begin{equation}
    \begin{split}
        f(r) &= \frac{(r^2-r_+^2)(r^2-r_-^2)}{r^2\ell_3^2}~,\qquad  N^\psi(r) = - \frac{r_+r_-}{r^2\ell_3}~.
    \end{split}
\end{equation}
The BTZ black hole has an outer and inner horizon at $r=r_\pm$. In terms of these parameters, the mass $M$ and angular momentum $J$ are 
\begin{equation}
    M = \frac{r_+^2 + r_-^2}{\ell_3^2}~, \qquad J = \frac{2r_+ r_-}{\ell_3}~,
\end{equation}
while the Hawking temperature $T_{\rm 3D}$ and angular velocity $\Omega_H$ read
\begin{equation}
    T_{\rm 3D} = \frac{r_+^2-r_-^2}{2\pi \ell_3 r_+}~,\qquad \Omega_H = \frac{r_-}{r_+}~.
\end{equation}
With this parameterisation, the BTZ black hole falls in a natural way in the decomposition \eqref{metric ansatz}, where it is straightforward to identify the background metric and dilaton as
\begin{equation}
   g_{ab} \dd x^a \dd x^b = - f(r) \dd t^2 + \frac{\dd r^2}{f(r)}~, \qquad \Phi(x)=r~.
\end{equation}
In this notation we also have that 
\begin{equation}
Q= J =\frac{2r_+r_-}{\ell_3}~.    
\end{equation}

At extremality, where $r_\pm = r_H$ and $T_{\rm 3D} =0$, the near-horizon geometry develops an AdS$_2$ throat.
Near-extremality is achieved by slightly turning on the temperature. This means that we parameterise how far away we are from extremality through the small parameter $\epsilon \ll 1$, where 
\begin{equation}\label{eq:rpm-decouple}
    r_\pm = r_H \pm \epsilon \, \xi~, \qquad  T_{\rm 3D} = \frac{\epsilon \, \xi}{ \pi \ell_2}+ \mathcal{O}(\epsilon^2)~.
\end{equation}
Note that $\xi$ is a near-extremality parameter which is fixed as we take $\epsilon \to 0$, and with some hindsight, we are using \eqref{eq:Phi0-3d} to trade $\ell_3$ for $\ell_2$. To zoom into the AdS$_2$ geometry, and be able to keep track of the backreaction, we define the new coordinate system $(\rho,\hat\tau,\phi)$ as 
\begin{equation}
    r = r_H + \epsilon \, \xi\,  \cosh\rho/\ell_2~, \qquad t = \ell_2 \frac{\hat \tau}{\epsilon}~, \qquad \psi = \phi + \frac{\hat\tau}{2\epsilon}~.
    \label{definition IR coordinates}
\end{equation}
The metric, to leading order in $\epsilon$, is
\begin{equation}
\begin{split}
 ds^2 &= \dd\rho^2 - \xi^2 \sinh^2(\rho/\ell_2) \dd\hat\tau^2 \\
     &\quad + r_H^2 \bigg(\dd\phi 
     + \frac{\xi}{2 r_H} \cosh(\rho/\ell_2)\dd\hat\tau\bigg)^2+ \mathcal{O}(\epsilon).
     \label{eq:leading-order-nhne-metric}
\end{split}
\end{equation}
The top line describes what we call an AdS$_2$ black hole, and the second line shows that the total space is a fibration of a circle over the near-AdS$_2$ spacetime. It is instructive to map this geometry to the parameters used for the AdS$_2$ solution in Sec.\,\ref{sec:general-2d-EA}. The background AdS$_2$ metric matches that in \eqref{eq:bh-bg}, where the parameter $\xi$ is related to the temperature via 
\begin{equation}\label{eq:beta-btz}
  \beta= \frac{2\pi \ell_2}{\xi}~.
\end{equation}
And we can also see that \eqref{eq:Phi0-3d} holds: $\Phi_0^2=r_H^2= Q\ell_2$.

We will be interested in the corrections to the metric at first order in $\epsilon$: this defines our backreacted AdS$_2$ background in \eqref{eq:near-AdS2} for BTZ, which dictates the couplings in our effective action in two dimensions.  The response to leading order in $\epsilon$ is
\begin{equation}
\begin{split}
    g_{ab} &= \bar{g}_{ab} + \epsilon\, h_{ab}+\cdots\,, \\
    \Phi(x) &= \Phi_0+ \epsilon\, \mathcal{Y}+\cdots\,.
    \end{split}
\end{equation}
Here $\bar{g}$ is the AdS$_2$ metric which can be read from \eqref{eq:leading-order-nhne-metric}. The responses of these quantities in $\epsilon$ are parameterised by $h$ and $\mathcal{Y}$, where the latter describes the correction to the size of the $U(1)$ circle. These quantities are easily obtained by keeping the first-order corrections in $\epsilon$ as we implement the decoupling limit \eqref{eq:rpm-decouple}-\eqref{definition IR coordinates}. This gives
\begin{equation}
    \begin{split}
        h_{\hat\tau \hat\tau} &= \frac{\xi^3}{r_H} \sinh(\rho/\ell_2 )^2 \cosh(\rho/\ell_2  )\,,\\
        h_{\rho \rho} &= \frac{\xi}{r_H}\cosh(\rho/\ell_2 )\,,\\
        \mathcal{Y} &= \xi\cosh(\rho/\ell_2)\,.
    \label{eq:BTZ corr}
    \end{split}
\end{equation}
In comparison with the notation in \eqref{TimeIndepDilaton}, we have that  $\mathcal{Y}_0 = \xi$. Note that the metric perturbation is symmetric and traceless as in \eqref{eq:gauge-hst} (and we can always do a diffeomorphism as described around \eqref{eq:h-hom}-\eqref{eq:int-scalar-gauge-2} to adjust for the trace terms).

\paragraph{Correction to the two-point function.}
So far, we have mapped the basic entries of the effective field theory of near-extremal BTZ to the two-dimensional notation used in Sec.\,\ref{sec:general-2d-EA}. We can now use the results in Sec.\,\ref{sec:corr-2pt-fn} to evaluate the two-point function of the massive field appearing in \eqref{eq:action-3d}. The effective action for $\varphi$ in the backreacted geometry of near-extremal BTZ is given by \eqref{I-massless-EFT} for massless fields and \eqref{I0-massive-EFT}  for massive ones, where the respective couplings are 
\begin{equation}
\begin{aligned}
    \lambda_0&= \frac{\epsilon}{2\Phi_0\ell_2^2}= \frac{\epsilon}{2r_H \ell_2^2}~,\\
    \lambda_{\mathfrak{m}} &= \epsilon\frac{\nu}{6}= \frac{\epsilon}{r_H}~.
\end{aligned}
\end{equation}
It is worth noting that with these values of the effective couplings, we have that the unified coupling in \eqref{def:c-delta} becomes
\begin{equation}\label{eq:c-delta-btz}
    c_\Delta= \frac{\Y_0}{\Phi_0}\epsilon = \frac{\xi}{r_H}\epsilon~, \qquad \forall~~ \Delta~,
\end{equation}
therefore we will have the same strength of the correction for both massive and massless fields. 
From \eqref{free piece} and \eqref{relationGcorrGfree}, we thus find that for integer $\Delta$, the Euclidean two-point function with first order correction reads
\begin{equation}\label{eq:Corr-BTZ-2d}
  \frac{ \langle {\cal O}_\Delta (n){\cal O}_\Delta(-n)\rangle_{\rm BTZ, 2D}}{ \langle {\cal O}_\Delta (n){\cal O}_\Delta(-n)\rangle_{\rm free}} = 1 +  \frac{\xi}{r_H} (1-2|n|H_{|n|})\epsilon +O(\epsilon^2)~.
\end{equation}
This is the expression we will contrast with the low-frequency/low-temperature limit of the correlation function computed via AdS$_3$/CFT$_2$.

\subsection{Two-point functions on the BTZ background}

Evaluating the two-point function on BTZ is a well-known result; see, for instance, \cite{BirSac02,SonSta02}. Following the effective action \eqref{eq:action-3d}, the basic idea is to solve the wave equation for a massive scalar on BTZ. If we decompose the field in Fourier components
\begin{equation}
    \varphi(t,r,\psi) = \sum_{k} \int\dd\omega \,e^{i \frac{\omega}{\ell_3}t - i k \psi}\, \varphi_{\omega,k}(r)~,
\end{equation}
then near the boundary of AdS$_3$ the field has a fall-off of the form
\begin{equation}
    \varphi_{\omega,k}(x) \underset{x\to\infty}{=} \psi_1(\omega,k) x^{\Delta-1}(1+\cdots)  + \psi_2(\omega,k) x^{-\Delta}(1+\cdots)~, \quad x \equiv \frac{r^2-r_+^2}{r_+^2}~. 
    \label{far solution BTZ}
\end{equation}
Note that here $\Delta$ is the conformal dimension as defined in AdS$_2$, i.e., the same as in \eqref{eq:Corr-BTZ-2d}; its relation to the conformal dimension in AdS$_3$ is simply
\begin{equation}
    \Delta_{\rm 3D}  = 2\Delta = 1+\sqrt{1+\mathfrak{m}^2\ell_3^2}~.
\end{equation}
By imposing ingoing boundary conditions at the horizon, we find explicit expressions for $\psi_1(\omega,k)$ and $\psi_2(\omega,k)$.
The retarded two-point function is, up to an overall normalisation, given by \cite{SonSta02}
\begin{equation}
    G^R_{\text{BTZ}}(\omega,k) = (2\Delta-1)\frac{\psi_2(\omega,k)}{\psi_1(\omega,k)}~.
\end{equation}
The explicit expression is 
\begin{equation}
    G^R_{\text{BTZ}}(\omega,k) = \tau_H^{2\Delta-1} (2\Delta-1)\frac{\Gamma(1-2\Delta)}{\Gamma(2\Delta-1)} \frac{\Gamma\left(\Delta - \frac{i(\sqrt{a}+\sqrt{b})}{\sqrt{\tau_H}}\right)\Gamma\left(\Delta + \frac{i(\sqrt{a}-\sqrt{b})}{\sqrt{\tau_H}}\right)}{\Gamma\left(1-\Delta - \frac{i(\sqrt{a}+\sqrt{b})}{\sqrt{\tau_H}}\right)\Gamma\left(1-\Delta + \frac{i(\sqrt{a}-\sqrt{b})}{\sqrt{\tau_H}}\right)}\,.
    \label{eq:exact-2pt}
\end{equation}
where 
\begin{equation}
 a(\omega,k) = \frac{\ell_3^2}{4\tau_H r_+^4}(r_-\omega-r_+k)^2~, \quad  b(\omega,k) = \frac{\ell_3^2}{4\tau_H r_+^4}(r_+\omega-r_-k)^2~,
\end{equation}
and
\begin{equation}
    \tau_H \equiv \frac{r_+^2-r_-^2}{r_+^2}~.
\end{equation}

It is important to note that we will be using the retarded Green's function in the following rather than the Euclidean two-point function. The reason is that in Euclidean signature, the periodicity in Euclidean time quantises the frequency $\omega$, which is inconvenient when taking a low-frequency limit. In contrast, we can easily take the low-frequency limit of the retarded correlator; the only minor subtlety is to reconstruct from it the Euclidean answer after taking the limit, which we address below.

\paragraph{Low-frequency/low-temperature limit.} The important step now is to implement the near-extremal limit on the retarded correlator \eqref{eq:exact-2pt}. The discussion is guided by the decoupling limit \eqref{eq:rpm-decouple}-\eqref{definition IR coordinates}; implementing this on the mode decomposition of the field gives  
\begin{equation}
\begin{aligned}\label{eq:decomp-field-near}
    \varphi(t,r,\psi)  &= \sum_k \int \dd\omega \;\varphi_{k,\omega}(r)\;e^{ i \frac{\omega}{\ell_3}t-i k \psi}\\
    &= \sum_k \int \dd\omega \;\varphi_{k,\omega}(r)\;e^{i \frac{\hat\tau}{2\epsilon}(\omega-k)-i k \phi  }~.
\end{aligned}
\end{equation}
From here, we identify the infrared Fourier variables as
\begin{equation}
    k_{\text{IR}} = k~, \quad \epsilon\, \omega_{\text{IR}} = \frac{1}{2}(\omega-k)~.
\end{equation}
We can see that in the $s$-wave sector (where $k=0$), taking $\epsilon\to0$ with $\omega_{\text{IR}}$ fixed is equivalent to having a low frequency $\omega$.
The two-point function \eqref{eq:exact-2pt}, expressed in IR coordinates, is given by
\begin{equation}
    G^R_{\text{BTZ}}(\omega_{\text{IR}},k_{\text{IR}}) =(2\Delta-1) \tau_H^{2\Delta-1} \frac{\Gamma(1-2\Delta)}{\Gamma(2\Delta-1)}\frac{\Gamma(\Delta-\frac{i\ell_2}{\xi}\omega_{\text{IR}})\Gamma(\Delta - \frac{i\ell_2}{r_H}k_{\text{IR}} - \frac{i\epsilon \ell_2}{r_H}\omega_{\text{IR}})}{\Gamma(1-\Delta-\frac{i\ell_2}{\xi}\omega_{\text{IR}})\Gamma(1-\Delta - \frac{i\ell_2}{r_H}k_{\text{IR}} - \frac{i\epsilon \ell_2}{r_H}\omega_{\text{IR}})}~.
\end{equation}
This is still an exact equation, where we made the dependence on $
\epsilon$ explicit in favour of $r_\pm$ and appropriate frequencies, but we have not implemented a near-extremal limit.  We can expand to first order in $\epsilon$, while keeping the remaining variables fixed: this takes $T_{\rm 3D}\to 0$ and $\omega\to 0$, and we find
\begin{align} \nonumber
   G^R_{\text{BTZ}}(\omega_{\text{IR}},0) 
    &=(2\Delta-1)\tau_H^{2\Delta-1} \frac{\Gamma(1-2\Delta)}{\Gamma(2\Delta-1)}\frac{\sin\left(\pi \left(\Delta + i \frac{\ell_2}{\xi}\omega_{\text{IR}}\right)\right)\sin(\pi \Delta)}{\pi^2} \bigg|\Gamma\left(\Delta-i\frac{\ell_2}{\xi}\omega_{\text{IR}}\right)\Gamma\left(\Delta \right)\bigg|^2\\
    &\qquad \left(1+i \pi\frac{\ell_2}{r_H}\omega_{\text{IR}} \cot(\pi \Delta)\,\epsilon \right)+ \mathcal{O}(\epsilon^2)~.
     \label{eq:s-wave-nh-2-pt}
\end{align}
 where we restricted to the $s$-wave sector by setting $k_{\text{IR}}=0$.

To compare with our analysis in Sec.\,\ref{sec:corr-2pt-fn}, which applied to BTZ is given in \eqref{eq:Corr-BTZ-2d}, we first have to construct the Euclidean two-point function from the retarded correlator, and then specialise to integer $\Delta$. Let us first discuss the construction of the Euclidean correlator. First, from the second line in \eqref{eq:decomp-field-near}, we expect a Wick rotation of the frequency, that is  
\begin{equation}
    \omega_{\text{IR}} = i |\omega_{\text{E}}|~.
\end{equation}
Next, the Euclidean frequency $\omega_{\rm E}$ is subject to the periodicity of Euclidean time: taking $\hat\tau= i\tau$, where $\tau\sim \tau+ \beta$, implies that the Euclidean frequencies are quantised as
\begin{equation}
    \omega_{\text{E}} = \frac{n\xi}{\ell_2} = \frac{2\pi n}{\beta }, \qquad n \in \mathbb{Z}~,
    \label{eq:discrete-momenta}
\end{equation}
with $\beta$ given in \eqref{eq:beta-btz}. With this, the Euclidean two-point function is related to the retarded propagator, in the $s$-wave sector, via \cite{SonSta02}
\begin{equation}
   \langle {\cal O}_\Delta(n){\cal O}_\Delta(-n)\rangle_{\rm 3D} = - G^R_{\text{BTZ}}\left(\omega_{\text{IR}}=i \frac{2\pi |n|}{\beta },0\right)
\end{equation}
From \eqref{eq:s-wave-nh-2-pt} we can then write
\begin{equation}\label{eq:low-T-2pt-BTZ}
         \langle {\cal O}_\Delta(\tau){\cal O}_{\Delta}(0)\rangle_{\rm 3D} =   \langle {\cal O}_\Delta(\tau){\cal O}_{\Delta}(0)\rangle_{\rm 3D,0} +  \langle {\cal O}_\Delta(\tau){\cal O}_{\Delta}(0)\rangle_{{\rm 3D},\epsilon} + O(\epsilon^2)
\end{equation}
where the leading order term is 
\begin{equation}\label{eq:Euc-2pt-BTZ-correc}
\langle \mathcal{O}_\Delta(n)\mathcal{O}_\Delta(-n)\rangle_{\text{3D,0}} =  - \left(\frac{\tau_H}{4}\right)^{2\Delta-1}\frac{2\pi}{\Gamma(\Delta-\frac12)^2 \cos(\pi \Delta)} \frac{\Gamma\left(\Delta + |n|\right)}{\Gamma\left(1-\Delta + |n| \right)} ~,
\end{equation}
and the first correction due to the near-extremality parameter $\epsilon$ reads
\begin{equation}\label{eq:corr-anyDelta-BTZ}
\langle \mathcal{O}_\Delta(n)\mathcal{O}_\Delta(-n)\rangle_{\text{3D},\epsilon} =-\epsilon\frac{2\pi^2 \ell_2}{r_H\beta} \cot(\pi \Delta) |n|  \langle \mathcal{O}_\Delta(n)\mathcal{O}_\Delta(-n)\rangle_{\text{3D,0}}~.
\end{equation}
This correction is what we would like to compare with the corrections predicted from the backreaction of near-AdS$_2$. 

To start, it is instructive to first focus on the leading order piece \eqref{eq:Euc-2pt-BTZ-correc} and contrast it to the derivations in Sec.\,\ref{sec:corr-2pt-fn}: this term should reproduce the free correlator on AdS$_2$.  Comparing the Fourier transform of \eqref{eq:2pt-free} with \eqref{eq:Euc-2pt-BTZ-correc} we find
\begin{equation}
    \frac{\langle \mathcal{O}_\Delta(n)\mathcal{O}_\Delta(-n)\rangle_{\text{3D,0}}}{\langle \mathcal{O}_\Delta(n)\mathcal{O}_\Delta(-n)\rangle_{\text{free}}} =  \frac{\beta}{\pi^2 \ell_2 } \left(\frac{2\epsilon}{r_H}\right)^{2\Delta-1}~.
\end{equation}
We see that, up to a normalisation, the leading order correction obtained from BTZ and the effective theory of AdS$_2$ agree, as expected and in agreement with \cite{GhoMax19}.   

Next, we focus on the correction in \eqref{eq:corr-anyDelta-BTZ} and how it compares to \eqref{eq:Corr-BTZ-2d} for integer values of $\Delta$. Since the correction in \eqref{eq:corr-anyDelta-BTZ} is clearly divergent for integer $\Delta$, the comparison requires regulating appropriately this expression. The procedure to treat the divergence is as follows. We take
\begin{equation}
\Delta=  \Delta_{\scaleto{\mathbb{Z}}{4pt}} + \delta~,  \qquad  \Delta_{\scaleto{\mathbb{Z}}{4pt}} \in \mathbb{Z}^+~,
\end{equation}
 and subsequently send $\delta \to 0$ in \eqref{eq:corr-anyDelta-BTZ}. Since the only divergence is coming from $\cot(\pi \Delta)$, anything that is divergent as $\delta \to 0$ is of order $\delta^{-1}$. This means that the correlation function is of the form 
\begin{equation}
\begin{split}
   \langle \mathcal{O}_\Delta(n)\mathcal{O}_\Delta(-n)\rangle_{\text{3D},\epsilon}\bigg \rvert_{\Delta = \Delta_{\scaleto{\mathbb{Z}}{4pt}}+\delta}& = \frac{F_1(n^2)}{\delta} + F_2(n^2)  +  \widetilde{G}(n) + \mathcal{O}(\delta)~,
      \label{eq:nonanalytic1}
    \end{split}
\end{equation}
where $F_1(n^2)$ and $F_2(n^2)$ are polynomials of $n^2$. In configuration space, any polynomial dependence in $n^2$ amounts to a contact-term contribution and can be disregarded at finite distance; see, e.g., App.~C of \cite{BalBer13} for a similar analysis. All the physical ingredients are in the non-analytic finite piece, which reads
\begin{equation}
\begin{aligned}
    \widetilde{G}(n)=  \epsilon\frac{4\pi^2 \ell_2}{r_H\beta}  \left(\frac{\tau_H}{4}\right)^{2\Delta_{\scaleto{\mathbb{Z}}{4pt}}-1}&\frac{|n| (-1)^{\Delta_{\scaleto{\mathbb{Z}}{4pt}}}}{\Gamma(\Delta_{\scaleto{\mathbb{Z}}{4pt}}-\frac12)^2} \frac{\Gamma\left(\Delta_{\scaleto{\mathbb{Z}}{4pt}} + |n|\right)}{\Gamma\left(1-\Delta_{\scaleto{\mathbb{Z}}{4pt}} + |n| \right)}\times \\ &\qquad\qquad\times \left(\psi^{(0)}\left(\Delta_{\scaleto{\mathbb{Z}}{4pt}}+n\right) + \psi^{(0)}\left(1-\Delta_{\scaleto{\mathbb{Z}}{4pt}}+n\right)\right)~. 
\end{aligned}
    \label{eq:Gtilde}
\end{equation}
Here $\psi^{(0)}(x)$ is the digamma function. We can rewrite these digamma functions into harmonic numbers $H_{|n|}$ via the identity
\begin{equation}
    \psi^{(0)}\left(\Delta_{\scaleto{\mathbb{Z}}{4pt}}+n\right) + \psi^{(0)}\left(1-\Delta_{\scaleto{\mathbb{Z}}{4pt}}+n\right) = 2H_{|n|} - \frac{1}{|n|} - \sum_{m=1}^{\Delta_{\scaleto{\mathbb{Z}}{4pt}}-1}\frac{2m}{n^2-m^2} -2\gamma_E~, 
    \label{eq:digammaharmonic}
\end{equation}
where $\gamma_E$ is the Euler--Mascheroni constant.  Removing again all analytic terms in $n^2$, we write
\begin{equation}
\begin{aligned}
     \langle \mathcal{O}_{\Delta_{\scaleto{\mathbb{Z}}{4pt}}}(n)\mathcal{O}_{\Delta_{\scaleto{\mathbb{Z}}{4pt}}}(-n)\rangle_{\text{3D},\epsilon} = &-\epsilon\frac{4\pi^2 \ell_2}{r_H\beta}  \left(\frac{\tau_H}{4}\right)^{2\Delta_{\scaleto{\mathbb{Z}}{4pt}}-1}\frac{(-1)^{\Delta_{\scaleto{\mathbb{Z}}{4pt}}}}{\Gamma(\Delta_{\scaleto{\mathbb{Z}}{4pt}}-\frac12)^2} \frac{\Gamma\left(\Delta_{\scaleto{\mathbb{Z}}{4pt}} + |n|\right)}{\Gamma\left(1-\Delta_{\scaleto{\mathbb{Z}}{4pt}} + |n| \right)} \left(1-2|n|H_{|n|}\right)\\&\qquad + \text{contact terms} ~.
      \label{eq:nonanalytic2}
    \end{aligned}
\end{equation}
We thus find that the renormalised expression for \eqref{eq:corr-anyDelta-BTZ} for integer conformal dimension is
\begin{equation}
\frac{\langle {\cal O}_{\Delta_{\scaleto{\mathbb{Z}}{4pt}}}(n){\cal O}_{\Delta_{\scaleto{\mathbb{Z}}{4pt}}}(-n)\rangle_{\rm 3D} }{\langle {\cal O}_{\Delta_{\scaleto{\mathbb{Z}}{4pt}}}(n){\cal O}_{\Delta_{\scaleto{\mathbb{Z}}{4pt}}}(-n)\rangle_{\rm 3D,0}}=  1+ \frac{\xi}{r_H}\,\epsilon \left(1-2|n|H_{|n|}\right) + \text{contact terms} + O(\epsilon^2)~,
    \label{eq:2pt-harmonic}
\end{equation}
where we used $\beta=2\pi \ell_2/\xi$. Comparing to the answer in \eqref{eq:Corr-BTZ-2d}, arising from the effective action near-AdS$_2$, we see that we have a precise match! This is a non-trivial confirmation that our analysis in Sec.\,\ref{sec:general-2d-EA} and Sec.\,\ref{sec:corr-2pt-fn} correctly constructed the effective theory and evaluated the imprint of the backreaction on two-point functions. 

\paragraph{A prediction for non-integer \texorpdfstring{$\Delta$}{Delta}.}
The low-frequency/low-temperature limit of the BTZ two-point function gives us an opening on what we should expect to be the effect of the backreaction of AdS$_2$ more broadly. The expressions in \eqref{eq:low-T-2pt-BTZ}-\eqref{eq:corr-anyDelta-BTZ} allow us to infer that the AdS$_2$ backreaction effects on two-point functions should generically take the form
\begin{equation}
\begin{split}
(I_{\text{eff}})_{\mathfrak{m}^2 \neq 0} &= - \frac{1}{2}\sum_{n \in \mathbb{Z}} \tilde{\varphi}_n \tilde{\varphi}_{-n}\left(\langle \mathcal{O}_\Delta(n)\mathcal{O}_\Delta(-n)\rangle_{\text{free}} +\lambda_{\mathfrak{m}}\mathcal{Y}_0 \langle \mathcal{O}_\Delta(n)\mathcal{O}_\Delta(-n)\rangle_\epsilon + O(\epsilon^2)\right)~,
\end{split}
\end{equation}
where the free piece is 
\begin{equation}\label{non-integer-prediction-1}
    \begin{split}
\langle \mathcal{O}_\Delta(n)\mathcal{O}_\Delta(-n)\rangle_{\text{free}} &=  \frac{-1}{\cos(\pi \Delta)}\frac{2\pi^2}{\Gamma(\Delta-\frac12)^2}\left(\frac{\pi \ell_2}{\beta}\right)^{2\Delta-2} \frac{\Gamma(\Delta+|n|)}{ \Gamma(1-\Delta+|n|)} ~,
    \end{split}
\end{equation}
and the leading order correction due to the backreaction is 
\begin{equation}\label{non-integer-prediction}
\langle \mathcal{O}_\Delta(n)\mathcal{O}_\Delta(-n)\rangle_\epsilon =-\pi |n| \cot(\pi \Delta) \langle \mathcal{O}_\Delta(n)\mathcal{O}_\Delta(-n)\rangle_{\text{free}}~.    
\end{equation}
We have not been able to reproduce this answer via the methods in Sec.\,\ref{sec:corr-2pt-fn}. Still, we expect this to be the correct answer for real $\Delta$.
 This agrees with the BTZ answer  if $\lambda_{\mathfrak{m}} = \frac{\epsilon}{r_H}$ and $\mathcal{Y}_0 = \xi $, and it agrees with our AdS$_2$ computations if we carefully take the limit of $\Delta \to \Delta_{\scaleto{\mathbb{Z}}{4pt}}$ as explained above.

We can Fourier transform back the expressions \eqref{non-integer-prediction-1}-\eqref{non-integer-prediction} to have the vertex correction for general $\Delta$. Following our definitions in  \eqref{eq:2pt-func} we recover the known result \eqref{eq:2pt-free} and obtain a correction of the form, 
\begin{equation} \label{eq:corr-explicit-general}
   \frac{\langle {\cal O}_\Delta (\tau){\cal O}_\Delta(0)\rangle}{\langle {\cal O}_\Delta (\tau){\cal O}_\Delta(0)\rangle_{\rm free}}= 1+\lambda_{\mathfrak{m}}\mathcal{Y}_0
    \frac{\pi ^{3/2}  \Delta  \csc (\pi 
   \Delta ) }{2^{\Delta}\Gamma \left(\Delta
   +\frac{1}{2}\right)} \sin ^{\Delta -1}\Big(\frac{\pi  \tau
       }{\beta }\Big)\Re\{P_{\Delta}^{\Delta -1}\big(i\cot\Big(\frac{\pi\tau}{\beta}\Big)\big)\} + O(\epsilon^2)~. 
\end{equation}
When $\Delta$ is an integer, this expression matches \eqref{Delta-integer-corrected}. 

\section{Higher-dimensional black holes} \label{sec:higher-d-exs}

In this section, we consider two-dimensional effective field theories that arise from the dimensional reduction of higher-dimensional ($D>3$) models: $\mathcal{N} = 2$ four-dimensional ungauged supergravity, and the five-dimensional Einstein--Hilbert action (with and without a negative cosmological constant). These effective field theories, and the associated corrections to the two-point functions of scalar fields present in these models, were previously studied in \cite{CasVer21,CasPed21}. Here, we revisit those results in light of the new analysis presented in Sec.\,\ref{sec:general-2d-EA} and Sec.\,\ref{sec:corr-2pt-fn}, and provide a comment on the physical interpretation of the correction to the two-point function. 

\subsection{Dyonic black holes in \texorpdfstring{${\cal N} = 2$}{N=2} 4D ungauged supergravity} \label{sec:dyonic-bhs}

Before specialising to a specific background, we briefly describe the generic supergravity theory studied in \cite{CasVer21}. The basic bosonic ingredients of this theory are the metric $g_{\mu\nu}^{(4)}$; six real scalar fields that are split into three dilatons, $\varphi_i$, and three axions, $\chi_i$, with $i = 1, 2, 3$; and four gauge fields $A^I$ with $I = 1,2,3,4$. For a more complete discussion of this theory, including the Lagrangian description, see \cite{CasVer21} and references therein. 

We consider configurations in four dimensions that respect spherical symmetry; the extremal near-horizon region then is precisely AdS$_2\times S^2$. This allows us to build an effective two-dimensional description in the spirit of Sec.\,\ref{sec:general-2d-EA} by integrating out the 2-sphere. We write these backgrounds as
    \begin{equation} \label{eq:4d2d}
       ds^4 = \frac{1}{\Phi(x)} g_{ab} \dd x^a \dd x^b + \Phi^2(x) \left(\dd \theta^2+\sin^2\theta\dd \phi^2\right).
    \end{equation}
We use $x^a$ to denote the two-dimensional coordinates, and $g_{ab}$ is the two-dimensional metric. The scalar $\Phi$ is the volume of the 2-sphere; it appears as a conformal factor for $g_{ab}$ to ensure that there is no kinetic term for $\Phi$ in the two-dimensional action. The matter field will also comply with spherical symmetry: there is also an appropriate ansatz for the field strengths (which can be magnetic or electric) supported by \eqref{eq:4d2d}, and the six scalar fields will only depend on $x^a$.

Specialising to the STU model, an example of ungauged ${\cal N}=2$ supergravity, and using the ansatz described above, leads to an effective two-dimensional description that fully captures the dynamics of four-dimensional spherically symmetric backgrounds. The final result is \cite{Lar18,CasVer21}
    \begin{equation} \label{eq:2d-action-sugra}
        S_{\rm 2D} = \frac{1}{4G_4}\int \dd^2x \sqrt{-g} {\cal L}_{\rm 2D}\,,
    \end{equation}
where
  \begin{equation} \label{eq:L2}
 \begin{aligned}
 {\cal L}_{\rm 2D}=&~\Phi^2 R^{(2)} + \frac{2}{\Phi} 
 -\frac{\Phi^2}{2}\sum_{i=1}^3 \left( (\partial_a \varphi_i)(\partial^a \varphi_i)  + e^{2\varphi_i} (\partial^a \chi_i )(\partial_a \chi_i)\right)- \frac{1}{2\Phi^3} U({\bf P},{\bf Q})~.
\end{aligned}
 \end{equation}
Note that as in Sec.\,\ref{sec:BTZ-BHs}, but contrary to the discussion in Sec.\,\ref{sec:general-2d-EA}, again $\Phi$ has units of length: this can be seen directly from \eqref{eq:4d2d}, where $\Phi(x)$ parameterises the size of the 2-sphere. Still, $\Phi^2/G_4$ is dimensionless, and as we discuss below, it is related to $\Phi/G_2$ in Sec.\,\ref{sec:general-2d-EA}. Here, $U({\bf P},{\bf Q})$ is a scalar potential encoding the magnetic and electric charges,
\begin{equation}\label{potential}
U({\bf P},{\bf Q})\equiv ({\bf P}^I ~~  {\bf Q}_I)
   \begin{pmatrix}
    (1+\chi_1^2 e^{2\varphi_1})k_{IJ} &- 2 e^{2\varphi_1}(kh)_I^{~\,J}\\
  - 2 e^{2\varphi_1} (hk)^{I}_{~J} & (k^{-1})^{IJ}\\
 \end{pmatrix}
   \begin{pmatrix}
 {\bf P}^J  \\
 {\bf Q}_J
 \end{pmatrix} ~.
\end{equation}
This Lagrangian is a consistent truncation for the $s$-wave sector of STU supergravity when compactified on $S^2$. The matrices $h_{IJ}$ and $k_{IJ}$ can be found in App.\,A of \cite{CasVer21}. The magnetic and electric charges $P^I$ and $Q_I$ are contained in the bold charges ${\bf P}^I$ and ${\bf Q}_I$, reflecting the dualization of the gauge fields:
  \begin{equation}
  {\bf Q}_I \equiv (Q_1,P^2, P^3,Q_4)~,\qquad {\bf P}^I \equiv (P^1,-Q_2,-Q_3,P^4)~.
  \end{equation}
From this action, one can obtain the equations of motion for the dilaton $\Phi$, the scalar fields $(\varphi_i, \chi_i)$ and the two-dimensional metric $g_{ab}$. They can be found in \cite{CasVer21}. 

We now further specialise to an example that will simplify the equations while retaining the relevant features: a dyonic 4D black hole with two electric and two magnetic charges turned on. Concretely, we will set $Q_3 = 0 = Q_4$ and $P^3=0=P^4$, and this gives 
  \begin{equation} \label{eq:dyonic2D}
 \begin{aligned}
 {\cal L}_{\rm 2D}=&~\Phi^2 R^{(2)} + \frac{2}{\Phi} -\frac{\Phi^2}{2}\sum_{i=1}^3 \left( (\partial_a \varphi_i)(\partial^a \varphi_i)  + e^{2\varphi_i} (\partial^a \chi_i )(\partial_a \chi_i)\right)- \frac{1}{2\Phi^3} U_{\rm dyonic}(\varphi_i, \chi_i)~,
\end{aligned}
 \end{equation}
where, for this choice of charges, the scalar potential \eqref{potential} now simplifies to
\begin{equation}
\begin{aligned} \label{eq:dyonic-potential}
    U_{\rm dyonic}(\varphi_i,\chi_i) =&~  e^{-\varphi_1+\varphi_2-\varphi_3} Q_2^2+ e^{\varphi_1-\varphi_2-\varphi_3} Q_1^2   +e^{\varphi_1+\varphi_2-\varphi_3}(  Q_2 \chi_1 + Q_1 \chi_2)^2
    \\
    &+ e^{\varphi_1-\varphi_2+\varphi_3}\big(P^2 -Q_1\chi_3\big)^2 + e^{-\varphi_1+\varphi_2+\varphi_3}(P^1-Q_2\chi_3)^2  \\ &+ e^{\varphi_1+\varphi_2-\varphi_3}\big(\chi_2(P^2-Q_1\chi_3)+\chi_1(P^1-Q_2\chi_3)\big)^2~.
\end{aligned}    
\end{equation}

\subsubsection{Effective theory near-AdS$_2$} \label{sec:dyonic-bg}
In what follows, we construct the effective theory that describes the AdS$_2$ backreaction for the 2D theory \eqref{eq:dyonic2D}-\eqref{eq:dyonic-potential}. In particular, we will discuss the AdS$_2$ background solution of interest, the linearised perturbations around it, and the interactions of the matter fields with the JT sector (which encodes the deviations away from extremality that are characteristic of near-AdS$_2$). We will match with the notation used in Sec.\,\ref{sec:general-2d-EA}.

\paragraph{AdS$_2$ background: dyonic non-BPS branch.}
As discussed previously, all AdS$_2$ backgrounds are characterised by having all of the scalars in play equal to a constant: this is the characteristic feature of an attractor mechanism. At the attractor point,
    \begin{equation}\label{eq:att-4d-bhs}
        \varphi_i = \bar{\varphi_i}~, \quad \chi_i = \bar{\chi_i}~, \quad \Phi = \Phi_0~, \quad g_{ab} = \Phi_0\, \bar{g}_{ab}\,,
    \end{equation}
where the right-hand sides represent constant values for the scalar fields. Notice that compared to \eqref{eq:lin-fluc}, we extracted a factor of $\Phi_0$ from the background metric; this compensates the powers in \eqref{eq:4d2d}. Studying the equations of motions at this fixed point reveals that the metric $\bar{g}_{ab}$ is locally AdS$_2$ with radius $\ell_2$, where 
\begin{equation}\label{eq:bg-dil-L2}
\ell_2^2 = \Phi_0^2~,
\end{equation}
implying that the AdS$_2$ and the $S^2$ radius are equal to each other. 

The attractor mechanism fixes the constant values of the scalars in \eqref{eq:att-4d-bhs}. For the potential \eqref{potential}, the attractor equations  give  $\barc_1 = 0 = \barc_2$, and further set 
    \begin{equation}\label{eq:att-dyonic}
    \begin{aligned}
         P^1 &= (c\, e^{-\barp_3}\pm \barc_3) e^{\barp_1-\barp_2}Q_1 \\
         P^2 &= (\mp c\, e^{-\barp_3}+\barc_3)Q_1 \\
         Q_2 &= \pm e^{\barp_1-\barp_2}Q_1~,       
    \end{aligned}
    \end{equation}
where $c \in  \{1,-1\}$, and  
    \begin{equation} \label{eq:bg-charge-l2}
        \ell_2^2 = e^{\barp_1-\barp_2-\barp_3}Q_1^2~.
    \end{equation}
Without loss of generality, we will pick $c = 1$ and the upper signs in \eqref{eq:att-dyonic}.

One can easily see that this solution is non-BPS by computing the Cayley hyperdeterminant \cite{KalKol96, Duff06}; see \cite{CasVer21} for details. Non-BPS solutions have negative Cayley hyperdeterminant $\hat{\bf{\Delta}}$, and setting $Q_{3,4} = 0 = P^{3,4}$ gives $\hat{\bf{\Delta}} = - \frac{1}{16}(P^1 Q_1 - P^2Q_2)^2$.

\paragraph{Effective action near-AdS$_2$.}
To single out the interactions between the dilaton and the scalar fields, we want to diagonalise the potential \eqref{eq:dyonic-potential} and suppress any interactions among the scalars. Guided by the attractor values \eqref{eq:att-4d-bhs}-\eqref{eq:att-dyonic}, we write 
    \begin{equation} \label{eq:probe-fields-4d}
        \varphi_i = \bar{\varphi}_i +  \bph_i ~, \qquad 
         \chi_i = \bar{\chi}_i + e^{-\bar{\varphi}_i}\bch_i~,
    \end{equation}
where we remind the reader that $\barc_{1,2}=0$ for the four-charge dyonic ungauged case. Following \cite{CasVer21}, we introduce
    \begin{equation} \label{eq:eigenfields}
    \begin{aligned}
        \mathfrak{Z}_1 &= \frac{1}{\sqrt{2}}\left( - \bph_1 + \bch_3 \right),\\
        \mathfrak{Z}_2 &=  \frac{1}{\sqrt{6}} \left( \bph_1 + 2\bph_2 +  \bch_3\right),\\
        \mathfrak{Z}_3 &= \frac{1}{\sqrt{3}} \left( \bph_1 - \bph_2 +\bch_3 \right)\,\\ 
        \mathfrak{Z}_{4} &= \bph_3\,,\quad \mathfrak{Z}_5 = \bch_1\,,\quad \mathfrak{Z}_6 = \bch_2~.
    \end{aligned}
    \end{equation}
On this basis, up to quadratic order in the fields, the potential $U({\bf{P,Q}})$  reduces to 
    \begin{equation}\label{eq:U-quadratic}
        U({\bf{P,Q}}) = 4\ell_2^2 +6 \ell_2^2 \mathfrak{Z}_3^2 + 2 \ell_2^2 \sum_{i = 4}^6 \mathfrak{Z}_i^2 + \cdots~,
    \end{equation}
where we used \eqref{eq:bg-charge-l2}. We see that the fields $\mathfrak{Z}_i$ diagonalise the quadratic fluctuations, 
where the fields $\mathfrak{Z}_{1,2}$ are massless, while the remaining fields are massive. Implementing \eqref{eq:att-dyonic} in the action \eqref{eq:dyonic2D}, and keeping only quadratic terms in the scalars, gives
    \begin{equation} \label{eq:2d-action-eigenfields}
 {\cal L}_{\rm 2D}=\Phi^2 R^{(2)} + V(\Phi)  -\frac{1}{2} \sum_{i=1}^6  \left(\Phi^2(\partial_a \mathfrak{Z}_i)(\partial^a \mathfrak{Z}_i)  + \frac{\ell_2^4}{\Phi^3} \mathfrak{m}_i^2 \mathfrak{Z}_i^2\right) +\cdots~,       
    \end{equation}
where we used \eqref{eq:U-quadratic}, and hence the dilaton potential is
    \begin{equation}
        V(\Phi) = \frac{2}{\Phi} - \frac{2\ell_2^2}{\Phi^3}~,
    \end{equation}
 and the masses of the scalar fields are 
    \begin{equation}\label{eq:masses-dyonic}
        \mathfrak{m}_{1,2}^2\ell_2^2 = 0~, \quad \mathfrak{m}_3^2\ell_2^2 = 6~, \quad \mathfrak{m}_{4,5,6}^2\ell_2^2 = 2~.
    \end{equation}

The action \eqref{eq:2d-action-eigenfields} is of the form \eqref{eq:gen-dil-model} up to some of the powers of $\Phi$: both in front of the Ricci scalar as well as in the kinetic and mass terms for the scalars. These differences are partly due to the choice of dimensional reduction \eqref{eq:4d2d}, and partly due to a more complicated coupling between the scalar fields and the field strengths in the original 4D action. To get to the precise form of \eqref{eq:gen-dil-model}, we could in principle rescale $\Phi^2 \to \Phi$; this would leave us only with a factor $\Phi_0^{-3/2}$ in front of the mass terms. To leading order, this will only change some of the constants in the effective action; the physics is completely similar to our discussion in Sec.\,\ref{sec:general-2d-EA}. We will therefore keep the Lagrangian in its current form to derive the effective action, which will lead to the same cubic effective actions as discussed in Sec.\,\ref{sec:corr-2pt-fn}. 

To extract the effects of near-AdS$_2$, we now linearise using
    \begin{equation}
        \Phi = \Phi_0 + \epsilon~  \Y \,,\qquad g_{ab} = \Phi_0\, \bar{g}_{ab} + \epsilon~  h_{ab}\,.
    \end{equation}
The equations of motion to linear order in $\epsilon$ are 
    \begin{equation} \label{eq:lin-eqs-4d}
    \begin{aligned}
        (\bn_a \bn_b - \bar{g}_{ab}\bar\square)\Y +\frac{1}{\ell_2^2} \bar{g}_{ab}\Y = 0~,\\
        \bar{R}^{ab} h_{ab} - \bn^a\bn^b h_{ab} + \bar\square h^a_{~a} + \frac{12}{\ell_2^2} \Y = 0~.
    \end{aligned}
   \end{equation}
Comparing to \eqref{eq:lin-dil}, we have $\nu = 12$.\footnote{It is instructive to match more precisely with the notation in Sec.\,\ref{sec:general-2d-EA}. If we redefine $\Phi_{\rm here}^2 = \Phi_{\rm 2D}$, then we have $\Y_{\rm 2D} = 2\epsilon \Phi_0 \Y_{\rm here} $. It is simple to check that with respect to $\Phi_{\rm 2D}$ and $\Y_{\rm 2D}$ we have that $\nu= - V''(\Phi_{\rm 2D})$ as expected from \eqref{eq:lin-dil}.} For the reasons discussed in Sec.\,\ref{sec:general-2d-EA}, the scalars $\mathfrak{Z}_i$ do not contribute to the backreaction of AdS$_2$ classically, and hence they are ommitted in \eqref{eq:lin-eqs-4d}. 

Finally, the effective action for the scalar fields, which accounts for the backreaction of AdS$_2$, reads
\begin{equation}
    S_{\rm eff} = \int d^2x \sqrt{-\bar{g}} \left({\cal L}_{\rm kin} + {\cal L}_{\rm int}\right)  + O(\epsilon^2)~,
\end{equation}
with
    \begin{equation} \label{eq:int-lag-4d}
    \begin{aligned}
        {\cal{L}}_{\rm kin} &= - \frac{1}{2}(\bar{\partial} \mathfrak{Z}_i)^2 - \frac{1}{2}\mathfrak{m}_i^2 \mathfrak{Z}_i^2 ~,\\ 
        {\cal L}_{\rm int} &= - \frac{\epsilon}{\Phi_0}\left( \Y (\bar{\partial }\mathfrak{Z}_i)^2 - \frac{3}{2} \mathfrak{m}^2_i \Y\mathfrak{Z}_i^2 + \frac{1}{4} h^{a}_{~ a}\big((\bar{\partial} \mathfrak{Z}_i)^2 + \mathfrak{m}_i^2 \mathfrak{Z}_i^2 \big)  - \frac{1}{2}h^{ab}\bn_a\mathfrak{Z}_i\bn_b\mathfrak{Z}_i \right),
    \end{aligned}
    \end{equation}
where $\mathfrak{m}_i^2$ is given in \eqref{eq:masses-dyonic}, and we are leaving the sum over $i$ implicit. In a slight abuse of notation, we also rescaled $\mathfrak{Z}_i \to \frac{\Phi_0}{\sqrt{4G_4}}\mathfrak{Z}_i$ to absorb the prefactor in \eqref{eq:2d-action-sugra} and account for the background value of the metric determinant. The result for ${\cal L}_{\rm int}$ agrees with \eqref{eq:int-scalar-Y} up to a power of $\Phi_0$ in the interaction terms between $h_{ab}$ and $\mathfrak{Z}_i$, which can be traced back to a factor of $\Phi_0$ that we extracted from the background metric in \eqref{eq:att-4d-bhs}; a factor of $2$ in front of the $\Y(\partial \mathfrak{Z}_i)^2$ term, which can be traced back to the quadratic power of $\Phi$ in the two-dimensional action \eqref{eq:dyonic2D}; and a factor $-\frac{3}{2}$ in front of the mass term, which is explained in the discussion below \eqref{eq:masses-dyonic}. This agrees with the cubic effective action found in \cite{CasVer21}.

Following our discussion in Sec.\,\ref{sec:gen-corr-2pt}, we can split the metric perturbation into a trace and symmetric traceless part, and use the remaining gauge freedom to eliminate two of these three degrees of freedom. The final interaction terms for massless and massive scalars then are
    \begin{equation}\label{eq:4D-massless-vertex}
        ({\cal L}_{\rm int})_{\mathfrak{m}^2=0} = -\epsilon\frac{1}{\ell_2^3} \Y \mathfrak{Z}_i^2 ~, \qquad i = 1,2~,
    \end{equation}
and
    \begin{equation} \label{eq:massive-4d-vertex}
        ({\cal L}_{\rm int})_{\mathfrak{m}^2\neq0} = -\epsilon\frac{2}{\ell_2}\Y \mathfrak{Z}_i\partial_\rho^2 \mathfrak{Z}_i~, \qquad i = 3,4,5, 6~.
    \end{equation}
Comparing to the notation in \eqref{I-massless-EFT} and \eqref{I-massive-EFT}, we read off $\lambda_0 = \epsilon /\ell_2^3$ and $\lambda_{\mathfrak{m}} = 2\epsilon/\ell_2$.

Finally, we can quantify how the two-point function of the dual operators to $\mathfrak{Z}_i$ is affected by the backreaction of AdS$_2$, where for the ranges of masses in \eqref{eq:masses-dyonic} the conformal dimensions are 
\begin{equation}\label{eq:delta-4d}
    \Delta_{1,2}=1 ~, \quad \Delta_3= 3~, \quad \Delta_{4,5,6}= 2~.
\end{equation}
The correlation functions of these operators will be of the form \eqref{Delta-integer-corrected} where the correction is controlled by $c_\Delta$ in \eqref{def:c-delta}. From the above equations, we have
    \begin{equation}
        c_\Delta = \frac{2\epsilon\Y_0}{\ell_2}~,
    \end{equation}
and
    \begin{equation}
        \frac{c_{\Delta =1}}{c_{\Delta > 1}} = \frac{2\ell_2^2\lambda_0 \Y_0}{\lambda_{\mathfrak{m}} \Y_0} =  1~.
    \end{equation}
Recall that for the BTZ black hole in \eqref{eq:c-delta-btz}, we also found that $c_{\Delta >1} = c_{\Delta = 1}$. Still, it is important to point out that there is no guarantee of having  $c_{\Delta >1} = c_{\Delta = 1}$ for a general model. The effective potentials $V(\Phi)$ in more complicated cases can spoil this coincidence.  

\paragraph{Connection to near-extremal black hole in 4D.} Until now, the emphasis has been on the near-AdS$_2$ perspective. Here, we will connect our analysis to the appropriate static black hole in the four-dimensional theory. 

From \eqref{eq:att-dyonic}-\eqref{eq:bg-charge-l2} it is clear how the AdS$_2$ radius is related to the electric and magnetic charges carried by the black hole. What remains is finding $\Y_0$, the near-extremal horizon value of the dilaton, which carries information about the near-extremality parameter of the black hole. This information can be extracted by inspecting the black hole geometry. In 4D ungauged supergravity, a static black hole solution is of the form \cite{ChoCom14}
    \begin{equation}\label{eq:4D-metric}
        ds^2 = - \frac{R(r)}{W(r)}\dd t^2 + \frac{W(r)}{R(r)}\dd r^2 + W(r) \dd  \Omega_2^2~.
    \end{equation}
Here $R(r)$ is a quadratic polynomial in $r$, and the roots of $R(r)$ define the inner ($r_-$) and outer ($r_+$) horizon, while $W(r)^2$ is a quartic polynomial in $r$ \cite{ChoCom14}.  The Bekenstein--Hawking entropy and Hawking temperature of the black hole read
\begin{equation}
    S_{\rm BH} = \frac{\pi}{G} W(r_+)~, \qquad T_{\rm 4D}= \frac{R'(r_+)}{4\pi W(r_+)}~.
\end{equation}
From \eqref{eq:4D-metric} it is clear that upon performing a dimensional reduction to 2D the dilaton will be $\Phi(r) = \sqrt{W(r)}$.

Next, we implement the near-extremal limit. This follows closely the analysis of the BTZ black hole in \eqref{eq:rpm-decouple}-\eqref{definition IR coordinates}. For the black hole \eqref{eq:4D-metric}, we have
\begin{equation}\label{eq:rpm4d}
    r_\pm =r_H \pm \epsilon\, \xi ~, \qquad T_{\rm 4D}= \frac{\epsilon \xi}{2\pi \ell_2^2}~,
\end{equation}
where we note that 
\begin{equation}\label{eq:4D-2D-ext}
    W(r_H)= \Phi_0^2= \ell_2^2 ~.
\end{equation}
To zoom into AdS$_2\times S^2$ starting from \eqref{eq:4D-metric}, we define
\begin{equation}
    r= r_H+ \epsilon\, \xi \cosh\rho/\ell_2 ~, \qquad t = \ell_2 \frac{\hat \tau}{\epsilon}~,
\end{equation}
which to leading order in $\epsilon$ gives
\begin{equation}
    ds^2  \underset{\epsilon\to0}{=} \dd\rho^2 - \xi^2\sinh^2(\rho/\ell_2)\dd \hat\tau^2  + \ell_2^2 \dd  \Omega_2^2 + O(\epsilon)~.
\end{equation}
The AdS$_2$ inverse temperature is $\beta=2\pi\ell_2/\xi$. 
This is the attractor background that controls the moduli in \eqref{eq:att-4d-bhs}-\eqref{eq:bg-dil-L2}. Finally, using the expressions in \cite{ChoCom14}, we read off the dilaton via
\begin{equation}
\begin{aligned}
    \epsilon \Y &= \Phi(r)- \Phi_0  \\
    &=\epsilon \frac{\xi}{2\ell_2} W'(r_H) \cosh\rho/\ell_2 + O(\epsilon^2)~,
\end{aligned}
\end{equation}
 Hence $\Y_0 = \frac{\xi}{2\ell_2} W'(r_H)$. To provide a physical meaning to $ W'(r_H)$, notice that expanding the entropy near-extremality gives
\begin{equation}
\begin{aligned}
    S_{\rm BH} &= \frac{\pi}{G} W(r_H) + \frac{\pi}{G} W'(r_H) \epsilon\,\xi + O(\epsilon^2) \\
    &= \frac{\pi}{G} \Phi_0^2 + \frac{\pi^2\ell_2^2}{G} W'(r_H)\,  T_{\rm 4D} + O(T_{\rm 4D}^2)~.
\end{aligned}
\end{equation}
Thus $W'(r_H)$ controls the so-called ``mass-gap'', or equivalently, the heat capacity to leading order in temperature.
Note that the derivation of $\Y_0$ only uses that the black hole geometry is of the form \eqref{eq:4D-metric}. Hence, our result in \eqref{eq:final-4d-2pt} holds for any static solution of 4D ungauged supergravity. 

A simple way to probe asymptotically flat black holes is by setting up a scattering process, where waves of the scalar fields scatter off the horizon of the black holes. In the low-frequency and low-temperature limit, the effective near-AdS$_2$ region evaluates the appropriate correlation functions of these fields, which in turn are related to a cross section via the optical theorem. We will elaborate more on this relation in Sec.\,\ref{sec:discussion}. For now, we report that our analysis predicts the finite-temperature corrections of the correlation functions.  Concretely, we find that the corrected two-point function of the operators ${\cal O}_{\Delta}$, the dual operators to the fields $\mathfrak{Z}_i$ in \eqref{eq:eigenfields}, reads
    \begin{equation}\label{eq:final-4d-2pt}
            \frac{\langle {\cal O}_\Delta(t_E){\cal O}_\Delta(0)\rangle}{\langle {\cal O}_\Delta(t_E){\cal O}_\Delta (0)\rangle_{\rm free}} =   1 +  \pi W'(r_H) T_{\rm 4D}\left(2 + \Delta  \pi \frac{1-2 T_{\rm 4D}\, t_E}{\tan(\pi T_{\rm 4D} \, t_E)}   +   S_{\Delta} (t_E) \right)  + O(T_{\rm 4D}^2)~,
    \end{equation}
where, for the choice of charges \eqref{eq:dyonic2D}, the conformal dimensions $\Delta$ are given in \eqref{eq:delta-4d}. Note that we are writing \eqref{eq:final-4d-2pt} in terms of the black hole metric \eqref{eq:4D-metric}, using Euclidean time ($t = i t_E$), and 
\begin{equation}\label{eq:2d-4d}
    \frac{\tau}{\beta}=  T_{\rm 4D}  \,t_E~, 
\end{equation}
which relates 2D variables on the right-hand side to 4D variables on the left-hand side. 
The explicit expressions for $S_{\Delta>1} (\tau)$ (recall $S_{\Delta = 1}(\tau) = 0$) can be found in \eqref{eq:explicit-S-Delta}, where in writing $S_{\Delta}(t_E)$ it is understood that we are using \eqref{eq:2d-4d}.

It is worth comparing  \eqref{eq:final-4d-2pt} with \cite{CasVer21}, which also discussed the effects of  $\Y$ on the dynamics of the scalar fields $\mathfrak{Z}_i$. The analysis of \cite{CasVer21} used what we now understand is an incorrect computation of this correction, treating the dilaton field as a $\Delta = -1$ operator to mimic its behaviour as a background field as prescribed in \cite{MalSta16}. For $\Delta >1$ this produces the wrong value of $c_{\Delta}$ and the functional dependence in $t_E$ in \eqref{eq:final-4d-2pt}; see the discussion in Sec.\,\ref{sec:massless-explicit-vertex} and App.\,\ref{App:Vertex}. With the analysis here, we have corrected the appropriate expressions.

In addition, a puzzle that arose in \cite{CasVer21} was the appearance of an extremal correlator \cite{DHoFre99}: na\"ively taking the three-point function between two massless scalar fields ($\Delta_{\mathfrak{Z}}=1$) and the dilaton as a $\Delta_\Y = -1$ operator lead to a divergence due to $d = \Delta_\Y +2\Delta_\mathfrak{Z}$.\footnote{Usually, extremal cubic correlators between three fields $\phi_{1,2,3}$ appear whenever $\Delta_1 = \Delta_2 + \Delta_3$; in this computation, $\Y$ would be in the $\Delta_-$ branch, and we have $d = \Delta_- + \Delta_2 +\Delta_3$.}  However, as discussed above, the procedure of \cite{MalSta16} is not the right one to compute the correction to the two-point function; indeed, as we have shown in Sec.\,\ref{sec:massless-explicit-vertex}, properly treating $\Y$ as a background field in the interaction \eqref{eq:4D-massless-vertex} yields the top line of \eqref{Delta1corrected} with $\lambda_0 = \epsilon/\ell_2^3$, and the final answer in \eqref{eq:final-4d-2pt} is finite.

\subsection{Five-dimensional rotating black holes} \label{sec:5d-bhs}

The second example we will consider here is related to five-dimensional rotating black holes with a single rotational parameter. The dimensional reduction of the five-dimensional theory to a two-dimensional dilaton gravity was performed in \cite{CasLar18}, which we summarise briefly here. Starting from the five-dimensional Einstein--Hilbert action with a negative cosmological constant, we dimensionally reduce to two dimensions using
    \begin{equation} \label{eq:5Dmetric}
        g_{\mu\nu}^{(5)} \dd x^\mu \dd x^\nu = e^{\psi+\chi} ds_{(2)}^2 + L^2 e^{-2\psi+\chi} d\Omega_2^2 + L^2 e^{-2\chi}(\sigma^3 + A)^2~.
    \end{equation}
Here $\psi$ and $\chi$ are two scalar fields, $A = A_a \dd x^a$ is a one-form, and $L$ is a scale introduced to keep the scalar fields dimensionless. The two-dimensional metric is
    \begin{equation}
        ds_{(2)}^2 = g_{ab} \dd x^a \dd x^b~, \qquad a,b=0,1~,
    \end{equation}
we define $d\Omega_2^2$ to be the round metric on the sphere, 
    \begin{equation}
        d\Omega_2^2 = \dd \theta^2 + \sin^2\theta \dd \phi^2  = (\sigma^1)^2 + (\sigma^2)^2 ~,
    \end{equation}
and the angular forms are
    \begin{equation}
    \begin{aligned}
        \sigma^1 &= -\sin\psi \dd\theta+\cos\psi\sin\theta \dd\phi\,,\\
        \sigma^2 &= \cos\psi \dd\theta+\sin\psi\sin\theta \dd\phi\,,\\
        \sigma^3 &= \dd \psi + \cos\theta\dd\phi~.
    \end{aligned}
    \end{equation}
We are considering the $s$-wave sector of the dimensional reduction, hence all fields depend only on $x^a$. After integrating out the field strength associated to $A$, the two-dimensional action is
    \begin{equation} \label{eq:5d-action-reduced}
         S_{\rm2D} = \frac{1}{2\kappa_2^2} \int\hspace{-0.1cm} \dd^2 x\sqrt{-g}\ e^{-2\psi}\Big( R-\frac{L^2Q^2}{2}e^{3\chi+5\psi}  + \frac{1}{2L^2}\Big(4e^{3\psi} - e^{5\psi-3\chi}\Big) + \frac{12}{\ell_5^2}e^{\psi+\chi} - \frac{3}{2}(\nabla\chi)^2\Big)~.
    \end{equation}
Here $R$ is the two-dimensional Ricci scalar, and $\psi$ plays the role of the dilaton: it controls the area of the squashed $S^3$ in the five-dimensional spacetime \eqref{eq:5Dmetric}. The constant $Q$ is an electric charge in 2D due to integrating out $A_a$, and from a five-dimensional perspective controls the angular momentum of \eqref{eq:5Dmetric}. The scalar field $\chi$ plays the role of a more traditional matter field, whose non-trivial profile reflects that the $S^3$ is squashed. In the following, we will extract the relevant information about the effective theory that captures the backreaction of near-AdS$_2$ and quantify the finite-temperature effects on the appropriate correlation functions.

\subsubsection{Effective theory \texorpdfstring{near-AdS$_2$}{near-AdS2}}

Despite appearances, the results in Sec.\,\ref{sec:general-2d-EA} apply to the model in \eqref{eq:5d-action-reduced}. Although the action \eqref{eq:5d-action-reduced} is not of the form of \eqref{eq:gen-dil-model}, the features that define the backreaction of AdS$_2$ are still the same and hence the effective theory near-AdS$_2$ falls in the same class. The simplest way to see this is by decoding the information from the equations of motion, as done in \cite{CasPed21}, and this will allow for a simple map with the results in Sec.\,\ref{sec:general-2d-EA}. We will find many simplifications when  $\ell_5 \to \infty$, and comment on this case separately.

\paragraph{AdS$_2$ background.}\label{sec:5d-bg}
 The two-dimensional equations of motion derived from \eqref{eq:5d-action-reduced} can be found in \cite{CasPed21}. Evaluating them at the attractor point (constant scalars) gives a locally AdS$_2$ metric with 
    \begin{equation}
        \frac{1}{\ell_2^2} = \frac{1}{2L^2}e^{3\psi_0}\left(-4 + 3e^{2\psi_0-3\chi_0}\right),
    \end{equation}
and the remaining equations of motion give
    \begin{equation}
        \frac{L^4Q^2}{2}e^{3\chi_0} = e^{-3\chi_0} - e^{-2\psi_0}~, \qquad \frac{1}{\ell_5^2} = \frac{1}{8L^2} e^{4\psi_0 -\chi_0}\left(e^{-3\chi_0} - 2e^{-2\psi_0}\right).
    \end{equation}
As was done in \cite{CasLar18,CasPed21}, we will trade the charge $Q$ for a non-negative constant $q$ defined through
    \begin{equation}
        q \equiv \frac{1}{8}e^{2\psi_0}(L^4Q^2e^{3\chi_0}-e^{-3\chi_0}) = \frac{1}{8} (e^{2\psi_0-3\chi_0}-2)\,.
    \end{equation}
Then 
    \begin{equation}\label{eq:L2-5d-bh}
        \frac{1}{\ell_2^2} = \frac{1}{L^2}e^{3\psi_0} (1+12q)\,, \qquad \frac{1}{\ell_5^2} = \frac{q}{L^2}e^{2\psi_0-\chi_0}\,.
    \end{equation}
We thus see that setting $q\to 0$ corresponds to setting the five-dimensional cosmological constant to zero (and $q\to\infty$ corresponds to a strongly coupled AdS$_5$ spacetime). 

\paragraph{Linear analysis.}
The equations of motion couple the perturbations around the scalar fields and the metric, and this is the technical obstruction that makes it cumbersome to compare \eqref{eq:gen-dil-model} with \eqref{eq:5d-action-reduced}. At the linearised level they can be decoupled by separating the perturbations into homogeneous and inhomogeneous parts, the latter of which will depend on $\Y$. Informed by the results in \cite{CasPed21}, we therefore linearise using
    \begin{equation} \label{eq:5d-linearise}
    \begin{aligned}
        e^{-2\psi} &= e^{-2\psi_0} + \epsilon~ \Y~, \\
        \chi &= \chi_0 + \epsilon\,\left( \frac{2q}{1+2q} e^{2\psi_0} \Y + \varphi \right), \\
        g_{ab} &= \bar{g}_{ab} + \epsilon\,\left( \bar{g}_{ab} \frac{6q}{1+2q} \varphi + h_{ab}\right).
    \end{aligned}
    \end{equation}
Here $(\psi_0, \chi_0, \bar{g}_{ab})$ define the $0$-th order background described above, and $(\Y,\varphi,h_{ab})$ the corresponding fluctuations. The linearised equations of motion for the dilaton field $\Y$ are then 
    \begin{equation}
        (\bn_a \bn_b - \bar{g}_{ab}\bar\square)\Y + \frac{1}{\ell_2^2} \bar{g}_{ab}\Y = 0\,,
    \end{equation}
as expected, and for the scalar field
    \begin{equation} \label{eq:KGeq-5d-scalar-field}
        \bar{\square} \varphi = \frac{1}{\ell_2^2} \frac{6+32q}{1+12q}\varphi~ \equiv \mathfrak{m}_\varphi^2 \varphi\, = \frac{\Delta(\Delta-1)}{\ell_2^2}\varphi~. 
    \end{equation}
Note that $ 6\leq \mathfrak{m}_\varphi^2 \ell_2^2 \leq 8/3$ (such that $2<\Delta \leq 3$) depending on the value of $q$. Finally, the equation of motion for the metric perturbation is   
    \begin{equation}
        \bar{R}^{ab}h_{ab} - \bn^a\bn^b h_{ab} + \bar{\square}h^a_{~a} + \frac{1}{\ell_2^2}\left( 6 e^{2\psi_0} \frac{1+10q+8q^2}{(1+2q)(1+12q)}\right)\Y = 0~,
    \end{equation}
from which we can read off
    \begin{equation}\label{eq:ch-5d-bh}
        \nu = 6 e^{2\psi_0} \frac{1+10q+8q^2}{(1+2q)(1+12q)}\,.
    \end{equation}
 With this, we see that the linear response is exactly of the same class as in  Sec.\,\ref{sec:gen-bg-eqs}.

\paragraph{Cubic effective action.}
The cubic effective action was also computed in \cite{CasPed21}; the relevant terms to our discussion are (after normalising the fields, and excluding self-interactions of $\varphi$)
    \begin{equation}
    \begin{aligned}
        {\cal L}_{\rm kin} &= - \frac{1}{2}(\bn \varphi)^2 - \frac{1}{2}\mathfrak{m}_\varphi^2\varphi^2~,\\
        {\cal L}_{\rm int} &= \epsilon \Big( \frac{e^{2\psi_0}}{2\ell_2^2}\frac{(1+6q)(9+38q-80q^2)}{(1+2q)^2(1+12q)}\Y \varphi^2- \frac{12q^2e^{2\psi_0}}{(1+2q)^2}\varphi(\bn\varphi)(\bn\Y) - \frac{1}{2} e^{2\psi_0}\Y(\bn\varphi)^2 \\ 
        &\qquad \quad- \frac{1}{4}h^{a}_{~a}\left((\bn\varphi)^2+m_\varphi^2\varphi^2\right)+\frac{1}{2} h^{ab}\bn_a\varphi\bn_b\varphi \Big)~.
    \end{aligned}
    \end{equation}
Compared to the discussion in Sec.\,\ref{sec:gen-cubic-action}, we have one additional term:  $\varphi(\bn\varphi)(\bn\Y)$. To leading order in $\epsilon$, this simply reduces to
    \begin{equation}
    \begin{aligned}
        \varphi(\bn\varphi)(\bn\Y) &= ({\rm tot. ~der.})- \frac{1}{2}\varphi^2 \bar{\square}\Y \\
        &= - \frac{1}{\ell_2^2}\varphi^2 \Y + O(\epsilon)~.
    \end{aligned}
    \end{equation}
As was the case for the 4D dyonic black hole in Sec.\,\ref{sec:dyonic-bhs}, the effective action is not an exact match with \eqref{eq:int-scalar-Y}: there is an additional constant in front of the $\Y\varphi^2$ term. This can be traced back to the coupling between $e^{\psi}$ and $\chi$ in \eqref{eq:5d-action-reduced}, reflected also in the inhomogeneous terms in \eqref{eq:5d-linearise}. However, since the field $\varphi$ is massive, the terms $\Y\varphi^2$ and $\Y(\bar\nabla\varphi)^2$ are total derivatives and can be discarded, as explained in Sec.\,\ref{sec:gen-corr-2pt}. We are left with 
    \begin{equation}
        {\cal L}_{\rm int} = \epsilon \frac{1}{2}h_{ab}^{\rm ST}\bn^a\varphi\bn^b\varphi~,
    \end{equation}
which in the black hole gauge \eqref{eq:bh-bg} reduces to 
    \begin{equation} \label{eq:5d-vertex}
        {\cal L}_{\rm int} =-\epsilon \frac{\nu}{6}\Y\varphi\partial_\rho^2 \varphi~,
    \end{equation}
where $\nu$ is given in \eqref{eq:ch-5d-bh}. 
    
\paragraph{Flat space limit.} \label{sec:5d-flat-limit}
In the regime $q=0$, where we have an asymptotically flat 5D spacetime, the action \eqref{eq:5d-action-reduced} simplifies and exactly fits our general model \eqref{eq:gen-dil-model} once we consider $\chi$ to be a probe field, which we can implement by setting
    \begin{equation}
        \chi = \chi_0 + \epsilon~ \varphi~.
    \end{equation}
Then 
    \begin{equation}\label{eq:flat-space-action}
        I_{\rm 2D} = \frac{1}{2\kappa_2^2}\int \dd^2x \sqrt{-g} e^{-2\psi} \left( R + \frac{2}{L^2}e^{3\psi} - \frac{e^{-3\chi_0}}{L^2} e^{5\psi} - \frac{3}{2}(\nabla \varphi)^2-\frac{9e^{-3\chi_0}}{2L^2} e^{5\psi}\varphi^2\right).
    \end{equation} 
We will rescale $\varphi$ such that its kinetic term is canonically normalised. Recognizing $\Phi = e^{-2\psi}$, we can now compare to \eqref{eq:gen-dil-model}. We can read off the potential  
    \begin{equation}
        V(\Phi) = \frac{2}{L^2}\Phi^{-1/2} - \frac{e^{-3\chi_0}}{L^2}\Phi^{-3/2}~. 
    \end{equation}
Due to the power of $\Phi$ and constant in front of the $\varphi^2$ term in \eqref{eq:flat-space-action}, the simple structure of the attractor equations is not as trivial as in Sec.\,\ref{sec:gen-bg-eqs}; in particular we do not find $\chi_0 = 0$, but instead
    \begin{equation}
        e^{3\chi_0} = \frac{1}{2}e^{2\psi_0}~.
    \end{equation}
The remaining equations in Sec.\,\ref{sec:gen-bg-eqs} do hold; in particular, we still find
    \begin{equation}
        V(\Phi_0) = 0 \quad \Leftrightarrow \quad \frac{1}{L^2} e^{\psi_0}\left(2-e^{2\psi_0-3\chi_0}\right)=0,
    \end{equation}
and
    \begin{equation}
        \bar{R}^{(2)} = - V'(\Phi_0) \quad \Leftrightarrow \quad \frac{1}{\ell_2^2}=\frac{1}{2L^2}e^{3\psi_0}\left(-1 + \frac{3}{2} e^{2\psi_0-3\chi_0}\right) = \frac{e^{3\psi_0}}{L^2}~.
    \end{equation}
Note that this agrees with \eqref{eq:L2-5d-bh} for $q=0$, as it should. 

The linear analysis of Sec.\,\ref{sec:gen-lin-analysis} yields 
    \begin{equation}
        \bar{\square}\varphi  
        = \frac{6}{\ell_2^2}\varphi~,
    \end{equation}
which agrees with \eqref{eq:KGeq-5d-scalar-field} for $q=0$. Notice that the mass (and corresponding conformal dimension) is an integer in this case ($\Delta = 3$). For the gravitational perturbation we find from \eqref{eq:ch-eq}
    \begin{equation}
        \frac{\nu}{\ell_2^2} =  -V''(\Phi_0) = \frac{6e^{2\psi_0}}{\ell_2^2}\,,
    \end{equation}
again agreeing with \eqref{eq:ch-5d-bh} for $q=0$. It is interesting to see that this theory collapses to the general, simpler action we studied in Sec.\,\ref{sec:general-2d-EA} only if the AdS radius is taken to infinity, which also gives the scalar field an integer mass. 

\paragraph{Backreaction effects on two-point functions.}
We now turn to our predictions for the corrected two-point function of the operator dual to $\varphi$. Recall that $\varphi$ is the fluctuation of $\chi$, which is a ``squashing" mode of the $S^3$ in the UV. 
In the flat space ($q=0$) limit, where $\Delta = 3$, we can use our explicit result for the two-point function correction for massive scalar fields \eqref{Delta-integer-corrected} to give a prediction for the corrected two-point function of the fluctuating squashing mode $\varphi$. For general values of $q$, the conformal dimension of the associated operator is not an integer, and our prediction for the corrected two-point function is \eqref{eq:corr-explicit-general}. For both cases, we need to find the value of $\Y_0$ to compute $c_\Delta = \lambda_\mathfrak{m}\Y_0$. The decoupling limit of the 5D rotating black hole was discussed thoroughly in \cite{CasLar18}, which gives
\begin{equation}
  \epsilon  \Y = e^{-2\psi}-e^{-2\psi_0} = e^{-2\psi_0}\frac{x(3-x^2)}{2(1+x^2)}\left(e^{\rho/\ell_2}+\varepsilon^2 e^{-\rho/\ell_2} \right) \frac{\lambda}{a_0}~,
\end{equation}
where $x = a_0/r_0$ with $a_0$ the extremal value of the rotational parameter and $r_0$ the horizon radius at extremality, $\varepsilon$ is the near-extremality parameter, and $\lambda$ is a dimensionful decoupling parameter introduced in \cite{CasLar18}; in relation to the BTZ expressions in \eqref{eq:rpm-decouple}, note that we have $2\varepsilon = \xi$. Comparing to the black hole gauge, where $\Y = \Y_0 \cosh(\rho/\ell_2)$, we can simply shift $\rho \to \rho - \ell_2\log\varepsilon$ such that 
\begin{equation}
\begin{aligned}
    \epsilon \Y_0 &= e^{-2\psi_0} \frac{x(3-x^2)}{(1+x^2)} \frac{\varepsilon\lambda}{a_0}\\
    &= e^{-2\psi_0}\frac{(1+2q)(1+8q)}{(1+4q)(1+6q)} \frac{r_0}{a_0^2}\varepsilon\lambda\,,
\end{aligned}
\end{equation}
where we further used $q = \frac{x^2-1}{4(2-x^2)}$. The AdS$_2$ temperature is then $\beta=\pi\ell_2/\varepsilon$. Thus we have 
    \begin{equation} \label{eq:cdelta-5d-general-q}
    \begin{aligned}
    c_\Delta&= \frac{\nu}{6}\Y_0\, \epsilon\\
    &=  \,\frac{(1+8q)(1+10q+8q^2)}{(1+4q)(1+6q)(1+12q)}\,\frac{r_0}{a_0^2}\varepsilon\lambda\\
    &= \frac{\pi  (5x^2-1-2x^4)}{(2x^2-1)^2} r_0\, T_{5\rm D} \,.
    \end{aligned}
    \end{equation}
Note that in the last line we wrote the correction in terms of $T_{\rm 5D}$, the Hawking temperature of the five-dimensional black hole. For general values of $q>0$, we have from \eqref{eq:corr-explicit-general}
    \begin{equation}\label{eq:ads5-correc}
    \begin{aligned}
        \frac{\langle {\cal O}_\Delta(\tau){\cal O}_\Delta(0)\rangle}{\langle {\cal O}_\Delta(\tau){\cal O}_\Delta (0)\rangle_{\rm free}} =   1 &+ c_\Delta  
    \frac{\pi ^{3/2}  \Delta  \csc (\pi 
   \Delta ) }{2^{\Delta}\Gamma \left(\Delta
   +\frac{1}{2}\right)} \sin ^{\Delta -1}\left(\frac{\pi  \tau
   }{\beta }\right)\Re\{P_{\Delta}^{\Delta -1}\left(i\cot \left(\frac{\pi  \tau }{\beta }\right)\right)\} \\ &\qquad + O(T_{\rm 5D}^2)~,
    \end{aligned}
    \end{equation}
with $c_\Delta$ as in \eqref{eq:cdelta-5d-general-q} and we remind the reader that $\Delta$ and $q$ are related via \eqref{eq:KGeq-5d-scalar-field}. In the flat space limit $q = 0$, we have
    \begin{equation}
 c_{\Delta = 3}= 2\pi r_0 \,T_{\rm 5D}~,
    \end{equation}
such that the corrected two-point function is, to subleading order,
    \begin{align}\label{eq:flat-correc} \nonumber
            \frac{\langle {\cal O}_3(\tau){\cal O}_3(0)\rangle}{\langle {\cal O}_3(\tau){\cal O}_3 (0)\rangle_{\rm free}} =   1&+  2\pi r_0 \, T_{\rm 5D}\Bigg(2 + 3   \frac{\pi-2\pi\tau/\beta}{\tan\Big(\frac{\pi \tau}{\beta}\Big)}   +   \frac{1}{10}\Big(27+14\cos\Big(\frac{2\pi\tau}{\beta}\Big) - \cos\Big(\frac{4\pi\tau}{\beta}\Big)\Big)\Bigg) \\ \qquad& + O(T_{\rm 5D}^2) ~,
    \end{align}
where we took $S_{\Delta = 3}(\tau)$ from \eqref{eq:explicit-S-Delta}. Note that we can easily express \eqref{eq:ads5-correc} and \eqref{eq:flat-correc} in terms of the UV time and temperature via the relation
\begin{equation}
    \frac{\tau}{\beta} = T_{\rm 5D} \, t_E~.
\end{equation}
This relation holds for both black holes that are asymptotically AdS$_5$ and flat. 

Finally, it is worth comparing \eqref{eq:ads5-correc} and \eqref{eq:flat-correc}  with the analysis in \cite{CasPed21}. First, we have corrected for the time dependence, with the results in this section being the correct expressions. Second, the method used in \cite{CasPed21}, which followed \cite{MalSta16}, led to a different value of $c_\Delta$. In particular, the value reported for $c_\Delta$ in \cite{CasPed21} did not have a definite sign, with a switch that depended on the value of $q$. This alternating sign was puzzling: from the point of view of the higher-dimensional UV theory, the two-point function correction---which is basically a correction to the density of states---should be positive definite. However, with the new analysis here, we find that, although the correction still depends on $q$ through the constant $c_{\Delta}$, it is positive definite for any value of $q$ as is clear from \eqref{eq:cdelta-5d-general-q}.

\section{Discussion} \label{sec:discussion}

Starting from a general dilaton gravity model coupled to a scalar field, we explored the imprint of the near-extremal gravitational backreaction on the dynamics of a scalar field. In particular, we evaluated the two-point function of the operator dual to the scalar field at tree-level. The effect of the backreaction of AdS$_2$ is to introduce a temperature correction in the correlation function.  

More concretely, in Sec.\,\ref{sec:general-2d-EA} we constructed the effective action that encodes this backreaction through cubic interactions between the scalar field and the near-AdS$_2$ sector (which consists of the dilaton field and the graviton). We found physically different results for massless versus massive scalar fields: massless scalar fields interact directly with the dilaton field, whereas for massive fields the interaction is mediated through the graviton. We encoded the imprint of the backreaction in the correction to the scalar two-point function as a result of these interactions. For integer conformal dimensions, we computed this correction exactly in Sec.\,\ref{sec:corr-2pt-fn} and found: 
    \begin{equation}\label{Delta-integer-corrected-discussion}
        \frac{\langle {\cal O}_\Delta(\tau){\cal O}_\Delta(0)\rangle}{\langle {\cal O}_\Delta(\tau){\cal O}_\Delta (0)\rangle_{\rm free}} =   1+ c_\Delta   \left(2 + \Delta  \pi \frac{1-2\tau/\beta}{\tan\big(\frac{\pi \tau}{\beta}\big)}   +   S_{\Delta} (\tau) \right)  + O(\epsilon^2)~,
    \end{equation}
where $S_{\Delta} (\tau)$ is given in \eqref{eq:S-Delta} and we note that $S_{\Delta=1} (\tau)$ vanishes, yielding a particularly clean expression for massless fields. The strength of the interaction is quantified by $c_{\Delta}$, given in \eqref{def:c-delta}, which is universal for massless fields but model-dependent for massive fields. Interestingly, in some of the simple examples that we study (e.g., the BTZ black hole and ungauged $\mathcal{N} = 2$ supergravity), $c_{\Delta}$ collapses to the same value for both massless and massive scalar fields. This ceases to be true for theories where there is a more complicated coupling between the dilaton and the scalar field. 

The dilaton gravity models that we consider are realistic effective descriptions of higher-dimensional black holes in the near-extremal, near-horizon limit. A well-controlled testing ground for our results is the BTZ black hole; in Sec.\,\ref{sec:BTZ-BHs}, we checked  \eqref{Delta-integer-corrected-discussion} against the subleading correction of the retarded two-point function of the operator dual to a scalar field on the BTZ black hole background. Not only were we able to provide an exact match with the subleading correction at low frequency and low temperature, confirming our result, but from this match we could also infer a prediction for the correction at non-integer conformal dimensions: 
    \begin{equation} \label{eq:corr-explicit-general-discussion}
       \frac{\langle {\cal O}_\Delta(\tau){\cal O}_\Delta(0)\rangle}{\langle {\cal O}_\Delta(\tau){\cal O}_\Delta (0)\rangle_{\rm free}} =   1+ c_\Delta 
        \frac{\pi ^{3/2}  \Delta  \csc (\pi 
       \Delta ) }{2^{\Delta}\Gamma \left(\Delta
       +\frac{1}{2}\right)} \sin ^{\Delta -1}\Big(\frac{\pi  \tau
       }{\beta }\Big)\Re\{P_{\Delta}^{\Delta -1}\big(i\cot\Big(\frac{\pi\tau}{\beta}\Big)\big)\}~.
    \end{equation}
This matches \eqref{Delta-integer-corrected-discussion} in the (subtle) integer-$\Delta$ limit. Finally, in Sec.\,\ref{sec:higher-d-exs}, we commented on several higher-dimensional settings: dyonic non-BPS black holes in ${\cal N}=2$, $D=4$ ungauged supergravity, and rotating five-dimensional black holes in AdS or flat space. We were able to resolve some open puzzles from \cite{CasVer21, CasPed21} related to the appearance of pathological extremal correlators and the sign of the correction, respectively. 

We end by elaborating on some potential applications and expansions of our results. 

\paragraph{Holographic dual interpretation: near-CFT$_1$.} It is worthwhile to interpret our results in the context of the near-AdS$_2$/near-CFT$_1$ correspondence: what might a microscopic dual description of the corrections we studied look like? Several aspects of JT gravity, which is a subset of the dilaton gravity models we studied here, are well described by a particular integrable limit of a class of SYK-like models, see, e.g., \cite{Sarosi:2017ykf} for a review. 
In \cite{Maldacena:2016hyu}, the correction to the two-point functions for a Majorana fermion in SYK was computed away from conformality at large $q$. That correction matches with \eqref{Delta-integer-corrected-discussion} for $\Delta=1$; however, the spectrum of the fermion ranges from $0<\Delta<1/2$. The answer reported in \cite{Maldacena:2016hyu} does not reproduce \eqref{eq:corr-explicit-general-discussion}. This creates a tension on how corrections to the conformal propagator in SYK should be contrasted with the corresponding expressions in gravity.

There are a couple of comments to make. First, as far as the near-AdS$_2$/near-CFT$_1$ correspondence is developed, there is no solid ground to expect that corrections to the SYK correlators should match those in gravity. Many aspects of SYK do not match classical gravity. Second, on a more positive (but speculative) note, it would be interesting to introduce Majorana fermions in our effective two-dimensional theory and evaluate the corrected two-point function. Given that our results are sensitive to $\Delta$, and likely also spin, this would be a more reliable test against the correction in SYK.  Finally, it would be interesting to investigate how to reproduce \eqref{eq:corr-explicit-general-discussion} from a near-CFT$_1$ perspective. This would shed light on what the conditions are for a holographic CFT at low energies to capture the appropriate dynamics of near-extremal black holes. We leave further considerations on this topic for future work.

\paragraph{Interplay with quantum corrections.} It is natural to ask how, if at all, the classical corrections we considered here compare to the quantum corrections studied in, e.g., \cite{BroIli24, Emp25, Big25,BetPap25}. Following initial work in \cite{BroIli24}, which studied the evaporation of charged black holes in the quantum regime, including the effects of near-extremal quantum backreaction on greybody factors of the emitted particles, \cite{Emp25, Big25} considered, among other things, the sensitivity of the two-point function of minimally coupled massless scalar fields in near-extremal (evaporating) black holes to quantum fluctuations in the throat. Such quantum corrections kick in when the energy of the black hole above extremality drops below a certain scale, typically denoted $E_{\rm brk}$, which signals a breakdown of the emergent near-horizon AdS$_2$ isometries. These corrections should be contrasted with the near-extremal \textit{classical} corrections that we computed here. As we lower the temperature of the black hole, the classical corrections will kick in before the quantum corrections; the latter will become far more important below $E_{\rm brk}$, and the effects of the coupling to matter fields generically become irrelevant. Although we do not expect an interplay between these two regimes, it would still be interesting to obtain a complete picture of the near-AdS$_2$ backreaction of matter fields as we follow the temperature all the way down to extremality. 

\paragraph{Quasinormal modes.} Our correction to the two-point function has a direct utility in the dynamical response of black holes at low temperatures. In particular, we can infer properties of quasinormal modes by inspecting the retarded two-point function we evaluated in near-AdS$_2$.  Writing our prediction for the retarded correlator in Fourier space, we have
\begin{equation}\label{non-integer-prediction-discussion}
G^{R}(\omega_{\rm IR}) = \Big(1 + \frac{i}{2} \lambda_{\mathfrak{m}} \Y_0\, \beta \,\omega_{\rm IR} \cot(\pi \Delta) \Big) G^{R}_{\text{free}}(\omega_{\rm IR})+\cdots~,    
\end{equation}
where we used the relation
\begin{equation}
   \langle {\cal O}_\Delta(n){\cal O}_\Delta(-n)\rangle = - G^R\left(\omega_{\text{IR}}=i \frac{2\pi |n|}{\beta }\right),
\end{equation}
and the free piece is
\begin{equation}\label{eq:free-fourier}
    \begin{split}
G^{R}_{\text{free}}(\omega_{\rm IR}) &=  \frac{1}{\cos(\pi \Delta)}\frac{2\pi^2}{\Gamma(\Delta-\frac12)^2}\left(\frac{\pi \ell_2}{\beta}\right)^{2\Delta-2} \frac{\Gamma(\Delta-i\frac{\beta \omega_{\rm IR}}{2\pi})}{ \Gamma(1-\Delta-i\frac{\beta \omega_{\rm IR}}{2\pi})} ~.
    \end{split}
\end{equation}
As a final step, we write this in terms of the black hole parameters. In the $s$-wave sector, we have
\begin{equation}
    \beta \omega_{\rm IR} = T_{\rm BH}^{-1} \omega~,
\end{equation}
with $T_{\rm BH}$ the Hawking temperature of the black hole measured by an asymptotic observer,  $\lambda_{\mathfrak{m}}\Y_0 \sim T_{\rm BH}$, and $\omega$ the frequency of time measured asymptotically. Hence \eqref{non-integer-prediction-discussion} reads
\begin{equation}\label{eq:BH-discussion}
G^{R}(\omega)\sim \Big(1 + i c \,  \omega \cot(\pi \Delta) \Big) \frac{\Gamma(\Delta-i\frac{\omega}{2\pi T_{\rm BH}})}{ \Gamma(1-\Delta-i\frac{\omega}{2\pi T_{\rm BH}})}+\cdots~,    
\end{equation}
with $c$ a temperature-independent constant, and we are ignoring overall normalisation factors. From this expression we can extract some useful information about the quasinormal modes of the black hole. At very low temperatures, the location is dictated by poles of the correlation function of AdS$_2$, i.e., the pole of the Gamma function in \eqref{eq:BH-discussion}; this is a well-known result \cite{Berti:2009kk}. The new result we see here is that our finite-temperature correction to the correlator \textit{does not} modify the location of these poles, it just affects the magnitude of the residue. This means that quasinormal modes encoded in this correlator do not receive an order $T^2_{BH}$ correction. Interestingly, this conclusion is universal regardless of the details of the black hole in question. 

Our effective theory only applies to neutral scalars, which limits the scope of what we can infer about quasinormal modes in the higher-dimensional theory. To expand the analysis, one first step is to incorporate a gauge field in the two-dimensional theory and have a matter field charged under it. From a higher-dimensional point of view, this would qualify as expanding the analysis beyond the $s$-wave sector, since the electric charge of the matter field would be playing the role of momenta. In this setup, it would be interesting to quantify the two-point function (including corrections) and study the behaviour of the quasinormal mode spectrum.  

\section*{Acknowledgments}
We thank Micha Berkooz, Roberto Empar\'an, Adam Levine, Thomas Mertens, Joan Sim\'on, Marija Toma\v{s}evi\'c, and Herman Verlinde for helpful discussions. AC has been partially supported by STFC consolidated grant ST/X000664/1. The work of JH is supported by the Dutch Black Hole Consortium with project number NWA.1292.19.202 of the research programme NWA which is (partly) financed by the Dutch Research Council (NWO). PJM was partially supported by CONICET and UNLP. EV has been supported by the Templeton Foundation via the Black Hole Initiative and the Gordon and Betty Moore Foundation, as well as by the ERC Consolidator grant (number: 101125449/acronym: QComplexity). Views and opinions expressed are however those of the authors only and do not necessarily reflect those of the European Union or the European Research Council. Neither the European Union nor the granting authority can be held responsible for them.

\appendix

 \section{Comments on interactions between dilaton and massive fields}\label{App:Vertex}

In this Appendix, we revisit a simple model of a non-derivative interaction between the dilaton and massive fields. We show by direct computation, drawing from similar tools as in \cite{FreMat98}, and in agreement with our results in Sec.\,\ref{sec:general-2d-EA}, that the interaction is trivial whenever $\Delta>1$. We comment on the differences with the approach used in \cite{MalSta16} and their results.

The simple model is
\begin{equation}\label{App-Vertex}
  I_{\rm eff}=\int \dd^2x\sqrt{\overbar{g}} \,\le(\frac{1}{2} \partial_a \varphi \partial^a \varphi + \frac{1}{2}\mathfrak{m}^2\varphi^2 +\lambda\, \Y \,\varphi^2 \ri)\,,
\end{equation}
where we will consider massive fields, with $\mathfrak{m}^2>0$, and we have introduced an effective coupling constant $\lambda$. In this model, the dilaton is taken to be a background field with the time-independent solution \eqref{TimeIndepDilaton} in an AdS$_2$ black hole background \eqref{eq:bh-bg}.   

To perform the bulk integrals in the vertex, it will be easier to go to Poincaré coordinates. To simplify the equations we will also take $\ell_2=1$ and $\beta=2\pi$ in the equations below. Using a standard change of coordinates, the classical solution for the dilaton becomes
\begin{equation}\label{eq:app-dilaton}
{\cal Y}= {\cal Y}_0 \cosh(\rho) = {\cal Y}_0 \frac{(z^2+x^2)+1}{2 z}~,
\end{equation}
and the position space bulk solution for massive scalar fields is
\begin{equation}\label{eq:SolPhi-0}
\begin{aligned}
\varphi_\Delta
&= \int\!\! \dd\tau' \;{\cal K}_{\Delta}(\rho;\tau,\tau') \,\tilde\varphi(\tau')= \int\!\! \dd x' \;{\cal K}_{\Delta}(z;x,x') \,\tilde\varphi(x') \\
&= \int\!\! \dd\tau' \left( \frac{\Gamma (\Delta )}{\sqrt{\pi }2^{\Delta}\, \Gamma
   \left(\Delta -\frac{1}{2}\right)} \frac{1}{(\cosh (\rho )-\sinh (\rho
   ) \cos (\tau -\tau' ))^{\Delta}}\right)\tilde\varphi(\tau')\\
&= \int\!\! \dd x' \left( \frac{\Gamma (\Delta )}{\sqrt{\pi }\, \Gamma
   \left(\Delta -\frac{1}{2}\right)} \left(\frac{z }{\left(x-x'\right)^2+z^2}\right)^{\Delta}\right)\tilde\varphi(x')  ~.
\end{aligned}
\end{equation}  
With these expressions at hand we write the on shell evaluation of the vertex as
\begin{align}
    I_{\rm int}
&=\lambda \int \dd^2x\sqrt{g} \,\varphi_\Delta\,\varphi_\Delta\, {\cal Y}\\ \nonumber
&=\lambda \int \left(\dd\rho \;\dd\tau \sinh(\rho)\right) \left(\int\!\! \dd\tau_1 \;{\cal K}_{\Delta}(\rho;\tau,\tau_1) \,\tilde\varphi(\tau_1) \right)\left(\int\!\! \dd\tau_2 \;{\cal K}_{\Delta}(\rho;\tau,\tau_2) \,\tilde\varphi(\tau_2) \right) {\cal Y}_0\cosh(\rho) \\ \nonumber
&=\lambda {\cal Y}_0 \int \left(\frac{\dd x \;\dd z}{z^2} \right) \left(\int\!\! \dd x_1 \;{\cal K}_{\Delta}(z;x,x_1) \,\tilde\varphi(x_1) \right)\left(\int\!\! \dd x_2 \;{\cal K}_{\Delta}(z;x,x_2) \,\tilde\varphi(x_2) \right) \left(\frac{(z^2+x^2)+1}{2 z}\right).
\end{align}
The time-independent dilaton profile imposes that the result of this integral must be a function of $(x_2-x_1)$. Hence, without loss of generality, we can just take $x_2=0$. The correction to the two-point function of ${\cal O}_{\Delta}$ due to the background field $\Y$ is
\begin{align}\label{Delta-1}
    \langle {\cal O}_\Delta(x_1){\cal O}_{\Delta}(0)\rangle_\lambda&= \lambda{\cal Y}_0\int \left(\frac{\dd x \;\dd z}{z^2} \right)\,{\cal K}_{\Delta}(z;x,x_1) \,{\cal K}_{\Delta}(z;x,0)\left(\frac{(z^2+x^2)+1}{z}\right) +O(\lambda^2)~.
\end{align}
We can now use inversion symmetry in the integrated variables, as in \cite{FreMat98}, by noticing that the solution ${\cal Y}$ is invariant under inversion. Under
\begin{equation}
    z\;\to\; \frac{\tilde z}{\tilde x^2+\tilde z^2}~,\qq\qq x\;\to\;\frac{\tilde x}{\tilde x^2+\tilde z^2}~,\qq\qq x_1\;\to\; \frac{1}{\tilde x_1}~,
\end{equation} 
we get
\begin{equation}
    \frac{(z^2+x^2)+1}{2z}\quad\to\quad \frac{(\tilde z^2+\tilde x^2)+1}{2\tilde z}\,,
\end{equation}
whilst
\begin{equation}
    \left(\frac{z}{\left(x-x_1\right)^2+z^2}\right)^{\Delta}
    \quad \to \quad
    |\tilde x_1|^{2\Delta} \left(\frac{\tilde z}{\left(\tilde x-\tilde x_1\right)^2+\tilde z^2}\right)^{\Delta}=\frac{1}{| x_1|^{2\Delta}} \left(\frac{\tilde z}{\left(\tilde x-\tilde x_1\right)^2+\tilde z^2}\right)^{\Delta}.
\end{equation}
Under these transformations, the integral becomes
\begin{align}\nn
    \langle {\cal O}(x_1){\cal O}(0)\rangle_\lambda & = \lambda{\cal Y}_0 \left(\frac{\Gamma (\Delta )}{\sqrt{\pi }\, \Gamma
   \left(\Delta -\frac{1}{2}\right)}\right)^2\frac{1}{|x_1|^{2\Delta}}\int_0^\infty \!\!\!\!\!\dd \tilde z\!\int_{\mathbb{R}} \!\!\!\dd \tilde x \frac{\tilde z^{2\Delta-2}}{(\tilde  z^2+(\tilde  x-\tilde x_1)^2)^{\Delta}}\left(\frac{(\tilde z^2+\tilde x^2)+1}{\tilde z}\right).
\end{align}
We now split the integral into two contributions and compute them independently as finite limits of some Feynman parameterised integrals. Notice also that all prefactors are regular for any value of $\Delta>1$, so in moving forward, we can drop all prefactors and focus on the result of the bulk integral alone. 
The two contributions are,
\begin{align}\label{INTA}
    A(\tilde x_1)&=\int_0^\infty \!\!\!\!\!\dd \tilde z\!\int_{\mathbb{R}} \!\!\!\dd \tilde x \frac{\tilde z^{2\Delta-2}}{(\tilde  z^2+(\tilde  x-\tilde x_1)^2)^{\Delta}}\left(\frac{(\tilde z^2+\tilde x^2)}{\tilde z}\right),\\
\label{INTB}
    B(\tilde x_1)&=\int_0^\infty \!\!\!\!\!\dd \tilde z\!\int_{\mathbb{R}} \!\!\!\dd \tilde x \frac{\tilde z^{2\Delta-2}}{(\tilde  z^2+(\tilde  x-\tilde x_1)^2)^{\Delta}}\left(\frac{1}{\tilde z}\right).
\end{align}
We now observe that these can be defined as regular instances of the integral \cite{FreMat98}
\begin{equation}\label{IntR}
I_{d+1}=\int_{0}^\infty \dd z_0 \int \dd^d\vec z \frac{z_0^a}{(z_0^2+(\vec z-\vec x)^2)^b(z_0^2+(\vec z-\vec y)^2)^c}\equiv I[a,b,c,d]|\vec x-\vec y|^{1+a+d-2b-2c}~,
\end{equation}
and
\begin{equation}\nn
I[a,b,c,d]=\frac{\pi^{d/2}}{2} \frac{\Gamma((a+1)/2)\Gamma(b+c-(d+a+1)/2)}{\Gamma(b)\Gamma(c)}\frac{\Gamma((1+a+d)/2-b)\Gamma((1+a+d)/2-c)}{\Gamma(1+a+d-b-c)}\;,
\end{equation}
where $A$ and $B$ can be thought of as cases of the above integral with $a=2\Delta-3$, $d=1$, $b=\Delta$ and in the limits $c=-1$ and $c=0$ respectively. We get
\begin{equation}\label{I-A11}
  I[2\Delta-3,\Delta,c,1]  =-\frac{\pi  }{(\Delta -1)
   \Gamma (c)}
   \frac{  \Gamma
   \left(c+\frac{1}{2}\right) \Gamma
   \left(-c+\Delta
   -\frac{1}{2}\right)}{\Gamma (-c+\Delta -1)}~,
\end{equation}
where we get that for our values of interest,
$$I[2\Delta-3,\Delta,-1,1]=I[2\Delta-3,\Delta,0,1]=0~,\qq  {\rm if} \quad \;\Delta>1\; ~.$$ 
Since both integrals $A$ and $B$ are smooth limits of the integral \eqref{IntR} independently, we find, for $\Delta>1$,
\begin{equation}
    \langle {\cal O}(x_1){\cal O}(0)\rangle_\lambda\sim \lim_{c\to-1}I[2\Delta-3,\Delta,c,1] + \lim_{c\to0}I[2\Delta-3,\Delta,c,1]=0~.
\end{equation}

Notice that a linear divergence appears in \eqref{I-A11} as $\Delta\to1$, potentially cancelling the $\Gamma^{-1}[c]$ zeroes and indicating a possible finite result for the vertex at $\Delta=1$. The limit is delicate to take in these coordinates using the arguments above, but we have already computed this vertex exactly in black hole coordinates in Sec.\,\ref{sec:corr-2pt-fn}.

Finally, we should comment on the correction reported in App.\,C of  \cite{MalSta16}. The integral in \eqref{INTA} is functionally what appears when evaluating a three-point function between two operators of conformal dimension $\Delta$ and a third one, $V_{-1}$, with conformal dimension ``$-1$''. Hence, the claim in \cite{MalSta16} is that the correction to the two-point function would be proportional to $\int \dd \tau \langle {\cal O}_\Delta {\cal O}_\Delta V_{-1}\rangle$. However, we note that 
\begin{equation}
   \int \dd \tau  \langle {\cal O}_\Delta {\cal O}_\Delta V_{-1}\rangle \sim \Gamma(-1)  A(\tilde x_1)~,
\end{equation}
where the extra Gamma function enters due to the normalisation of the bulk-to-boundary propagator of $V_{-1}$ if we use \eqref{eq:SolPhi-0} for this operator. This choice would make  $\int \dd \tau \langle {\cal O}_\Delta {\cal O}_\Delta V_{-1}\rangle$ finite, and up to a coefficient not specified, reproduces the result  App.~C of  \cite{MalSta16}. But this is clearly not the correct boundary condition, since the operator $V_{-1}$ should mimic \eqref{eq:app-dilaton} rather than \eqref{eq:SolPhi-0}. Therefore, what we have shown in this Appendix is that the appropriate normalisation of  $V_{-1}$ yields that the contribution of the interaction $\Y\varphi^2$ to the two-point function of ${\cal O}_\Delta$ is trivial.

\section{Solving the recurrence relation} \label{app:rec-rel}

In this Appendix, we explain how to derive the explicit result \eqref{eq:corr-explicit} for the correction to the two-point function from the recurrence relations in configuration space
\begin{equation} 
\langle {\cal O}_{\Delta+1} (\tau){\cal O}_{\Delta+1}(0)\rangle_{\rm free}
=\frac{4 \pi ^2 \ell_2^2 }{\beta
^2 (2 \Delta-1 )^2}
 \left(\Delta^2
   +\left(\frac{\beta}{2\pi }\partial_{\tau}\right)^2 \right) \langle {\cal O}_\Delta (\tau){\cal O}_\Delta(0)\rangle_{\rm free}~,
\end{equation}
where 
\begin{equation}
\langle {\cal O}_\Delta (\tau){\cal O}_\Delta(0)\rangle_{\rm free} \equiv \frac{1}{\beta^2}\sum_{n\in \mathbb{Z}}e^{-i\frac{2\pi n}{\beta}\tau}\langle {\cal O}_\Delta (n){\cal O}_\Delta(-n)\rangle_{\rm free}~.
\end{equation}
Together with \eqref{relationGcorrGfree} this implies,
\begin{equation} \label{eq:rec-rel-pos-space}
\langle {\cal O}_{\Delta+1} (\tau){\cal O}_{\Delta+1}(0)\rangle_{\epsilon}
=\frac{4 \pi ^2 \ell_2^2 }{\beta
^2 (2 \Delta -1)^2}
 \left(\Delta^2
   +\left(\frac{\beta}{2\pi }\partial_{\tau}\right)^2 \right) \langle {\cal O}_\Delta (\tau){\cal O}_\Delta(0)\rangle_{\epsilon}\,,
\end{equation}
where 
    \begin{equation}
\langle {\cal O}_\Delta (\tau){\cal O}_\Delta(0)\rangle_{\epsilon}\equiv \frac{1}{\beta^2}\sum_{n\in \mathbb{Z}}e^{-i\frac{2\pi n}{\beta}\tau}\langle {\cal O}_\Delta (n){\cal O}_\Delta(-n)\rangle_{\epsilon}~.
\end{equation}
Defining $\theta = \pi \tau/\beta$ for notational simplicity, we can rewrite \eqref{eq:rec-rel-pos-space} as
    \begin{equation} \label{eq:def-D-Delta}
    \begin{aligned}
        \langle {\cal O}_{\Delta+1} (\theta){\cal O}_{\Delta+1}(0)\rangle_{\epsilon} &= \left(\frac{2\pi\ell_2}{\beta}\right)^{2\Delta} \left(\prod_{k=0}^{\Delta -1} \frac{1}{(2\Delta - 1 -2k)^2}\left( (\Delta - k)^2 +\frac{1}{4}\partial_\theta^2\right) \right) \langle {\cal O}_{1} (\theta){\cal O}_{1}(0)\rangle_{\epsilon} \\
        &\equiv  \left(\frac{\pi\ell_2}{\beta}\right)^{2\Delta} \frac{\pi}{\Gamma\big(\Delta+\frac{1}{2}\big)^2}    D_{\Delta}  \left[\langle {\cal O}_{1} (\theta){\cal O}_{1}(0)\rangle_{\epsilon}\right]\,,
    \end{aligned}
    \end{equation}
and the same equation holds for $\langle {\cal O}_{\Delta+1} (\theta){\cal O}_{\Delta+1}(0)\rangle_{\rm free} $. Here we used
    \begin{equation}
        \left(\frac{2\pi\ell_2}{\beta}\right)^{2\Delta}\prod_{k=0}^{\Delta-1} \frac{1}{(2\Delta-1-2k)^2} = \left(\frac{\pi\ell_2}{\beta}\right)^{2\Delta} \frac{\pi}{\Gamma\big(\Delta+\frac{1}{2}\big)^2}\,.
    \end{equation}
Now we write (see \eqref{Delta1corrected} and \eqref{2pf-O0}) 
    \begin{equation} \label{eq:first-corr}
    \begin{aligned}
        \langle {\cal O}_{1} (\theta){\cal O}_{1}(0)\rangle_{\epsilon} &= \langle {\cal O}_{1} (\theta){\cal O}_{1}(0)\rangle_{\rm free} \left( 2 + \frac{\pi - 2\theta}{\tan\theta} \right) \\
        &= 2 \langle {\cal O}_{1} (\theta){\cal O}_{1}(0)\rangle_{\rm free} - (\pi - 2\theta) \frac{1}{2}\partial_\theta \langle {\cal O}_{1} (\theta){\cal O}_{1}(0)\rangle_{\rm free}~,
    \end{aligned}
    \end{equation}
such that
    \begin{equation}
    \begin{aligned}
        \langle {\cal O}_{\Delta+1} (\theta){\cal O}_{\Delta+1}(0)\rangle_{\epsilon}  =&~ 2\langle {\cal O}_{\Delta+1} (\theta){\cal O}_{\Delta+1}(0)\rangle_{\rm free}  \\
        &\quad- \frac{1}{2}\left(\frac{\pi\ell_2}{\beta}\right)^{2\Delta} \frac{\pi}{\Gamma\big(\Delta+\frac{1}{2}\big)^2}  D_\Delta\left[(\pi - 2\theta)\partial_\theta \langle {\cal O}_{1} (\theta){\cal O}_{1}(0)\rangle_{\rm free}\right],
    \end{aligned}
    \end{equation}
where the derivative operator $D_\Delta$ can be read from \eqref{eq:def-D-Delta} and expanded as
    \begin{equation}
        D_\Delta = \sum_{n=0}^{\Delta} a_n^\Delta \partial_\theta^{2n} = a_0^\Delta + \sum_{n=1}^{\Delta} a_n^\Delta \partial_\theta^{2n}~.
    \end{equation}
Notice that
    \begin{equation}
        \partial_\theta^{2n} \left[(\pi-2\theta)\partial_\theta \langle {\cal O}_{1} (\theta){\cal O}_{1}(0)\rangle_{\rm free} \right] = (\pi-2\theta) \partial_\theta^{2n+1}\langle {\cal O}_{1} (\theta){\cal O}_{1}(0)\rangle_{\rm free} - 4n \partial_\theta^{2n}\langle {\cal O}_{1} (\theta){\cal O}_{1}(0)\rangle_{\rm free}\,.
    \end{equation}
Therefore
    \begin{equation}
    \begin{aligned}
        D_\Delta\left[(\pi - 2\theta)\partial_\theta \langle {\cal O}_{1} (\theta){\cal O}_{1}(0)\rangle_{\rm free}\right] =& (\pi - 2\theta)\partial_\theta D_\Delta\left[\langle {\cal O}_{1} (\theta){\cal O}_{1}(0)\rangle_{\rm free}\right] \\
        &~ - 4\sum_{n=1}^\Delta n \,a_n^\Delta \partial_\theta^{2n} \langle {\cal O}_{1} (\theta){\cal O}_{1}(0)\rangle_{\rm free}~.
    \end{aligned}
    \end{equation}
Thus, we have 
    \begin{equation}
    \begin{aligned}
        \langle {\cal O}_{\Delta+1} (\theta){\cal O}_{\Delta+1}(0)\rangle_{\epsilon}  &= 2 \langle {\cal O}_{\Delta+1} (\theta){\cal O}_{\Delta+1}(0)\rangle_{\rm free}  - \frac{1}{2}(\pi-2\theta)\partial_\theta \langle {\cal O}_{\Delta+1} (\theta){\cal O}_{\Delta+1}(0)\rangle_{\rm free}  \\
        &\qquad \qquad +\left(\frac{\pi\ell_2}{\beta}\right)^{2\Delta} \frac{2\pi}{\Gamma\big(\Delta+\frac{1}{2}\big)^2} \sum_{n=1}^\Delta n\,a_n^\Delta  \partial_\theta^{2n} \langle {\cal O}_{1} (\theta){\cal O}_{1}(0)\rangle_{\rm free}\\ 
        &= \left(2 + (\Delta+1)\frac{\pi-2\theta}{\tan(\theta)}\right) \langle {\cal O}_{\Delta+1} (\theta){\cal O}_{\Delta+1}(0)\rangle_{\rm free} \\
        &\qquad \qquad +\left(\frac{\pi\ell_2}{\beta}\right)^{2\Delta} \frac{2\pi}{\Gamma\big(\Delta+\frac{1}{2}\big)^2} \sum_{n=1}^\Delta n\,a_n^\Delta  \partial_\theta^{2n} \langle {\cal O}_{1} (\theta){\cal O}_{1}(0)\rangle_{\rm free} ~.
        \end{aligned}
    \end{equation}
The coefficients $a_n^\Delta$ are given by
    \begin{equation}
        a_n^\Delta = \frac{\Gamma(\Delta+1)^2}{2^{2n}} \sum_{0\leq i_1<i_2<\ldots<i_n\leq \Delta-1} \frac{1}{(\Delta-i_1)^2(\Delta-i_2)^2\cdots(\Delta-i_n)^2}~,
    \end{equation}
where we used
    \begin{equation}
        \prod_{k=0}^{\Delta-1} (\Delta-k)^2 = \Gamma(\Delta+1)^2~.
    \end{equation}
In terms of unordered sums, 
    \begin{equation}
    \begin{aligned}
        a_n^\Delta = \frac{\Gamma(\Delta+1)^2}{2^{2n}n!} \sum_{i_1\neq i_2 \neq \ldots \neq i_n}^{\Delta-1} \frac{1}{(\Delta-i_1)^2(\Delta-i_2)^2\cdots(\Delta-i_n)^2}~.
    \end{aligned}
    \end{equation}
For example, for $n = 3$ this is
    \begin{equation}
    \begin{aligned}
       &\frac{a_3^\Delta}{\Gamma(\Delta+1)^2} = \frac{1}{2^6} \frac{1}{3!}\Big\{ \left( \sum_{i=0}^{\Delta-1}\frac{1}{(\Delta-i)^2} \right)^3\hspace{-0.2cm} - \binom{3}{2} 
        \sum_{i\neq j}^{\Delta-1} \frac{1}{(\Delta -i)^4(\Delta-j)^2} - \sum_{i=0}^{\Delta-1} \frac{1}{(\Delta-i)^6}\Big\} \\ 
        &= \frac{1}{2^6} \frac{1}{3!}\Big\{ \left( \sum_{i=0}^{\Delta-1}\frac{1}{(\Delta-i)^2} \right)^3 \hspace{-0.2cm}- 3\left( \sum_{i,j=0}^{\Delta-1}\frac{1}{(\Delta -i)^4(\Delta-j)^2} -\sum_{i=0}^{\Delta-1} \frac{1}{(\Delta-i)^6}\right)- \sum_{i=0}^{\Delta-1} \frac{1}{(\Delta-i)^6}\Big\} \\ 
        &= \frac{1}{2^6}\frac{1}{3!} \Big\{ \left( \sum_{i=0}^{\Delta-1}\frac{1}{(\Delta-i)^2} \right)^3 \hspace{-0.2cm}- 3\sum_{i,j=0}^{\Delta-1}\frac{1}{(\Delta -i)^4(\Delta-j)^2} +2 \sum_{i=0}^{\Delta-1} \frac{1}{(\Delta-i)^6}\Big\}~,
    \end{aligned}
    \end{equation}
which can be explicitly evaluated for general $\Delta$ using
    \begin{equation}
        \sum_{i=0}^{\Delta-1}\frac{1}{(\Delta -i)^n} = \sum_{j=1}^\Delta \frac{1}{j^n} = H_\Delta^{(n)} = \zeta(n) - \zeta(n,\Delta+1)~.
    \end{equation}
However, the combinatorics quickly become very cumbersome. There is an alternative, compact way of writing the coefficients $a_n^\Delta$ in terms of so-called elementary symmetric polynomials $e_n$. They are defined as
    \begin{equation} \label{eq:sym-pol-def}
        e_n(X_1, X_2, \ldots, X_k) = \sum_{1 \leq i_1 \leq \ldots i_n\leq k} X_{i_1}X_{i_2}\cdots X_{i_n}~.
    \end{equation}
Relabeling $j = \Delta - i$, we have $X_{j} =1/j^2$, such that
    \begin{equation}
        a_n^\Delta = \frac{\Gamma(\Delta+1)^2}{2^{2n}} \sum_{1\leq j_1<j_2<\ldots<j_n\leq \Delta} \frac{1}{j_1^2j_2^2\cdots j_n^2} = \frac{\Gamma(\Delta+1)^2}{2^{2n}}\ e_n\left(\frac{1}{1^2},\frac{1}{2^2},\frac{1}{3^2}, \ldots, \frac{1}{\Delta^2}\right).
    \end{equation}
What remains is to evaluate 
    \begin{equation}
        \tilde{S}_\Delta (\theta) \equiv \sum_{n=1}^{\Delta}\frac{n}{2^{2n}}\,e_n\left(1,\frac{1}{4},\frac{1}{9}, \ldots, \frac{1}{\Delta^2}\right)\,\partial_\theta^{2n}\csc(\theta)^2. 
    \end{equation}
We can use $\csc(\theta)^2 = - \partial_\theta \cot\theta$, and the identity
\begin{equation}
\begin{aligned}
\partial_\theta^m \cot\theta &= \delta_m \cot\theta -\delta_{m-1}\csc^2\theta \\
&\quad - m \sum_{k=0}^m\sum_{j=0}^{k-1} \frac{(-1)^j2^{m-2k}(k-j)^{m-1}}{(k+1)\sin(\theta)^{2+2k}} \binom{m-1}{k}\binom{2k}{j}
\sin\left(\frac{\pi m}{2} + 2(k-j)\theta\right).
\end{aligned}
\end{equation}
For $m=2n+1$ the first two terms vanish, and we get
    \begin{equation}
    \begin{aligned}
        \tilde{S}_{\Delta} (\theta) =& \sum_{n=1}^\Delta 2n(2n+1)(-1)^n e_n\Big(1,\frac{1}{4},\frac{1}{9}, \ldots, \frac{1}{\Delta^2}\Big)\times \\
        &\times \sum_{k=0}^{2n} \sum_{j=0}^{k-1} \frac{(-1)^j(k-j)^{2n}}{(k+1)2^{2k}\sin(\theta)^{2k+2}}\binom{2n}{k}\binom{2k}{j}  
        \cos(2(k-j)\theta)~.
    \end{aligned}
    \end{equation}
The final result is 
    \begin{equation}
    \begin{aligned}
        \langle {\cal O}_{\Delta+1} (\theta){\cal O}_{\Delta+1}(0)\rangle_{\epsilon} =& \left(2 + (\Delta+1)\frac{\pi-2\theta}{\tan(\theta)}\right) \langle {\cal O}_{\Delta+1} (\theta){\cal O}_{\Delta+1}(0)\rangle_{\rm free}  \\
        &~ + \frac{2\pi^2}{\beta^2}\left(\frac{\pi\ell_2}{\beta}\right)^{2\Delta} \frac{\Gamma(\Delta+1)^2}{\Gamma\big(\Delta+\frac{1}{2}\big)^2}  \tilde{S}_\Delta (\theta)\,,
    \end{aligned}
    \end{equation}
where  
    \begin{equation}
        \langle {\cal O}_{\Delta+1} (\theta){\cal O}_{\Delta+1}(0)\rangle_{\rm free} = \frac{\pi^2}{\beta^2}\left(\frac{\pi\ell_2}{\beta}\right)^{2\Delta} \frac{(2\Delta+1)\Gamma(\Delta+1)}{\sqrt{\pi}\Gamma(\Delta+\frac{1}{2})} \frac{1}{\sin(\theta)^{2\Delta+2}}\,.
    \end{equation}
We can now rewrite this to arrive at \eqref{eq:corr-explicit}, which we repeat here: 
    \begin{equation}
        \langle {\cal O}_{\Delta} (\tau){\cal O}_{\Delta}(0)\rangle_{\epsilon} = \langle {\cal O}_{\Delta} (\tau){\cal O}_{\Delta}(0)\rangle_{\rm free} \left(2 + \Delta\frac{\pi-2\pi\tau/\beta}{\tan(\frac{\pi \tau}{\beta})}   +  S_{\Delta} (\tau)
        \right)~,
    \end{equation}
where we restored $\tau=\beta \theta/\pi$ and defined
    \begin{equation}
   \begin{aligned}
        S_{\Delta} (\tau) \equiv & \, \frac{2\sqrt{\pi}\Gamma(\Delta)}{(2\Delta-1)\Gamma(\Delta-\frac{1}{2})} \sum_{n=1}^{\Delta-1} 2n(2n+1)(-1)^n e_n\Big(1,\frac{1}{4},\frac{1}{9}, \ldots, \frac{1}{(\Delta-1)^2}\Big) \times \\
        &\times \sum_{k=0}^{2n} \sum_{j=0}^{k-1} \frac{(-1)^j(k-j)^{2n}}{(k+1)2^{2k}\sin(\frac{\pi\tau}{\beta})^{2k+2-2\Delta}}\binom{2n}{k}\binom{2k}{j} 
        \cos\big(2(k-j)\frac{\pi\tau}{\beta}\big)\,.
    \end{aligned}        
    \end{equation}
We end by noting that for purposes of quick evaluation, the following rewriting for $S_{\Delta}(\tau)$ is more convenient: 
    \begin{equation}
        S_\Delta (\tau) =\frac{2\sqrt{\pi}\Gamma(\Delta)\sin(\frac{\pi\tau}{\beta})^{2\Delta}}{(2\Delta-1)\Gamma(\Delta -\frac{1}{2})}\sum_{n=1}^{\Delta-1}n \left(\frac{\beta}{2\pi}\right)^{2n}\,e_n\Big(1,\frac{1}{4},\frac{1}{9}, \ldots, \frac{1}{(\Delta-1)^2}\Big)\partial_\tau^{2n}\csc(\frac{\pi\tau}{\beta})^2~. 
    \end{equation}
Explicitly, the first few sums are
    \begin{equation}\label{eq:explicit-S-Delta}
    \begin{aligned}
        S_2(\tau) &= \frac{2}{3}\left(2+\cos(\frac{2\pi\tau}{\beta})\right)~,\\
        S_3(\tau) &= \frac{1}{10}\left(27+14\cos(\frac{2\pi\tau}{\beta}) - \cos(\frac{4\pi\tau}{\beta})\right)~,\\
        S_4(\tau) &= \frac{2}{105}\left( 214+113\cos(\frac{2\pi\tau}{\beta}) -13 \cos(\frac{4\pi\tau}{\beta}) + \cos(\frac{6\pi\tau}{\beta}) \right)~,\\
        S_5(\tau) &= \frac{1}{252}\left( 1375+734\cos(\frac{2\pi\tau}{\beta}) - 106\cos(\frac{4\pi\tau}{\beta})+14 \cos(\frac{6\pi\tau}{\beta}) - \cos(\frac{8\pi\tau}{\beta}) \right)~.
    \end{aligned}
    \end{equation}

\bibliography{all}{}
\bibliographystyle{ytphys}

\end{document}